\documentclass[usenatbib]{mn2e}
\usepackage{natbib}

\usepackage{amsmath}

\usepackage{epsfig}
\usepackage{graphicx}
\usepackage{amssymb}
\usepackage{aas_macros}
\usepackage{color}

\newcommand{\kms}{\,{\rm km \, s^{-1}}}
\newcommand{\kpc}{\,{\rm kpc}}
\newcommand{\mpc}{\,{\rm Mpc}}
\newcommand{\oversim}[2]{\protect{\mbox{\lower0.5ex\vbox{%
   \baselineskip=0pt\lineskip=0.2ex
   \ialign{$\mathsurround=0pt #1\hfil##\hfil$\crcr#2\crcr\sim\crcr}}}}} 

\catcode`\"=\active\let"=\"  

\def\3{{\ss} }

\def\c12{{1\over 2}}

\def\d{{\rm d}}   
   
\def\plusplus{\raise 0.3ex\hbox{${\scriptstyle ++}$}{}}

\bibliography{biblio}

\begin{document}   
\title[Modelling the local cosmic expansion]{A dynamical model of the local cosmic expansion}
\author[Jorge Pe\~{n}arrubia et al.]{Jorge Pe\~{n}arrubia$^{1,2,3}$\thanks{jorpega@roe.ac.uk}, Yin-Zhe Ma$^{4,5}$, Matthew G. Walker$^{6,7}$, Alan McConnachie$^{8}$\\
$^1$Institute for Astronomy, University of Edinburgh, Royal Observatory, Blackford Hill, Edinburgh EH9 3HJ, UK\\
$^2$Ram\'on y Cajal Fellow, Instituto de Astrof\'isica de Andalucia-CSIC, Glorieta de la Astronom\'ia s/n, 18008, Granada, Spain\\
$^3$Institute of Astronomy, University of Cambridge, Madingley Road, Cambridge CB3 0HA, UK\\
$^4$Department of Physics and Astronomy, University of British Columbia, Vancouver, V6T 1Z1, BC, Canada\\
$^5$Canadian Institute for Theoretical Astrophysics, Toronto, Canada\\
$^6$McWilliams Center for Cosmology\\
$^7$Department of Physics, Carnegie Mellon University, Pittsburgh, PA 15213, US\\
$^8$NRC Herzberg Institute of Astrophysics, 5071 West Saanich Road, Victoria, BC, V9E 2E7, Canada\\
}
\maketitle

\begin{abstract}
We combine the equations of motion that govern the dynamics of galaxies in the local volume with Bayesian techniques in order to fit orbits to published distances and velocities of galaxies within 3~Mpc. We find a Local Group (LG) mass $2.3\pm 0.7\times 10^{12}{\rm M}_\odot$ that is consistent with the combined dynamical masses of M31 and the Milky Way, and a mass ratio $0.54^{+0.23}_{-0.17}$ that rules out models where our Galaxy is more massive than M31 with $\sim 95\%$ confidence. The Milky Way's circular velocity at the solar radius is relatively high, $245\pm 23\kms$, which helps to reconcile the mass derived from the local Hubble flow with the larger value suggested by the `timing argument'. Adopting {\it Planck}'s bounds on $\Omega_\Lambda$ yields a (local) Hubble constant $H_0=67\pm 5\kms\mpc^{-1}$ which is consistent with the value found on cosmological scales. Restricted N-body experiments show that substructures tend to fall onto the LG along the Milky Way-M31 axis, where the quadrupole attraction is maximum. Tests against mock data indicate that neglecting this effect slightly overestimates the LG mass without biasing the rest of model parameters. We also show that both the time-dependence of the LG potential and the cosmological constant have little impact on the observed local Hubble flow. 
\end{abstract}   
\begin{keywords}
Galaxy: kinematics and dynamics 
\end{keywords}

\section{Introduction} \label{sec:dyn}
The dynamics of galaxies in and around the Local Group provide strong support in favour of expansionary cosmological theories. Nowhere else in the Universe can we map in such detail the competition between the primordial expansionary velocities and the gravitational pull of galaxies. To appreciate the beauty of this contest it is worth adopting a Friedmann-Lema\^itre-Robertson-Walker (FLRW) model wherein the distribution of matter in the Universe is isotropic and homogeneous and gravity behaves as predicted by the theory of General Relativity. Under these assumptions the relative motion between two mass-less particles (say A and G) in a flat Universe can be described by the Friedmann equations (e.g. Peacock 1999)
\begin{equation}\label{eq:friedmann}
\frac{\ddot r}{r}=-\frac{4 \pi G}{3}\rho + \frac{\Lambda c^2}{3};
\end{equation}
where $r=|{\bf r}_{\rm A}-{\bf r}_{\rm G}|$, $\rho$ is the density of the Universe, $c$ is the speed of light and $\Lambda$ is the cosmological constant. Defining the Hubble constant $H_0=(8\pi G \rho/3)^{1/2}$ and the fractional vacuum energy density at $z=0$ as $\Omega_{\Lambda}=\Lambda c^2/(3H_0^2)$ it is straightforward to show that the current constraints on these parameters, $H_0\approx 70\kms\mpc^{-1}$ and $\Omega_{\Lambda}\approx 0.7$, lead to a cosmological model in which the particles A and G must move away from each other regardless of the time at which they are observed. However, this is in stark contradiction with the large number of nearby galaxies that show blue-shifted spectra. The ingredient missing in Equation~(\ref{eq:friedmann}) is, of course, the gravity of the Local Group members.

Kahn \& Woltjer (1959) realized that this equation can be modified in order to measure the Local Group mass ($M$), resulting in what is typically known as the `timing argument'. Under the assumption that the mass in the Local Group is approximately made up by our Galaxy and Andromeda (M31) so that $M\approx M_{\rm G}+M_{\rm A}$, Equation~(\ref{eq:friedmann}) can be re-written as
\begin{equation}\label{eq:kep}
{\ddot r}=-\frac{G M}{r^2} + H_0^2\Omega_{\Lambda}r.
\end{equation}
Dropping the term with the cosmological constant in Equation~(\ref{eq:kep}) reduces the relative motion between the particles A and G to a Keplerian (radial) orbit. If the age of the Universe is known, the mass $M$ can be directly measured from the current separation and relative velocity between these galaxies. The original value obtained by Kahn \& Woltjer ($M\ge 1.8\times 10^{12}{\rm M}_\odot$) was considerably larger than the combined masses estimated from rotation curves of both galaxies ($\sim 2\times 10^{11}{\rm M}_\odot$), suggesting the presence of large amounts of ``unseen intergalactic matter'', a result which holds true after five decades of active research\footnote{At the time the distance between our Galaxy and M31 was thought to be $d\sim 600\kpc$, approximately $3/4$ of the current value (McConnachie 2012).}. 

Whether or not the cosmological constant plays an active role in the formation of the Local Group remains a matter of debate. For example, the velocity dispersion of nearby ($<3\mpc$) galaxies about the Hubble flow is much lower than predicted by cold dark matter simulations (Governato et al. 1997; Macci\`o et al. 2005; Karachentsev et al. 2008). Given the strong suppression of random motions in a fast expanding Universe (Baryshev et al. 2001; Chermin et al. 2001; Chermin 2004), this observation has been interpreted as a manifestation of vacuum energy on local scales (Teerikorpi et al. 2005). Such interpretation has been challenged by Hoffman et al. (2008), Peirani (2010) and Martinez-Vaquero et al. (2009), who find a marginal effect of the cosmological constant on local dynamics in N-body simulations with and without vacuum energy. Partridge et al. (2013) reach a similar conclusion through an analysis of the timing argument. 

The {\it local} Hubble flow may be fairly sensitive to the density environment in the vicinity of the Local Group. For example, Aragon-Calvo et al.(2011) find that the velocity fluctuations of galaxies embedded in large-scale structures tend to be smaller than around field galaxies. In this scenario the location of the Milky Way within a vast `wall' of structures connected to the Virgo Cluster (Tully \& Fisher 1987) may be responsible for the `coldness' of the local volume. Interestingly, these models also predict an enhanced value for the Hubble constant ($\simeq 77-113\kms\mpc^{-1}$) on $\sim 3\mpc$ scales. On the other hand, Wojtak et al. (2013) notice that the Hubble constant measured by observers embedded in low-density regions of the Universe tends to be systematically higher than the cosmological value. 

Despite the neat theoretical background on which Equation~(\ref{eq:kep}) rests, there remains a number of open questions regarding the dynamics of galaxies in the Local Group. For example, the mass returned from the timing argument when applied to the relative motion between the Milky Way and M31 ($M\sim 5\times 10^{12}{\rm M}_\odot$; Li \& White 2008; van der Marel et al. 2012a,b; Partridge et al. 2013 and references therein) is considerably larger than the combined dynamical masses of all galaxies in the Local Group ($\sim 2\times 10^{12}{\rm M}_\odot$). This result suggests the striking possibility that approximately half of the Local Group mass is missing. Van der Marel et al. (2012b) have inspected this long-standing issue in detail. First, using HST data these authors find that the transverse velocity of M31 is statistically negligible, supporting the assumption of radial motion in Equation~(\ref{eq:kep}). 
In addition, their analysis also reveals a strong dependence between the Local Group mass derived from the timing argument and the azimuthal velocity of the Sun within the Milky Way plane. Unfortunately, both the solar distance to the Galaxy centre and the local circular velocity remain unsettled (e.g. Sch\"onrich 2012 and references therein). 

There is also mounting evidence that the orbits of satellite galaxies around the Milky Way and Andromeda deviate from the distribution of substructures found in numerical simulations of structure formation. In particular, both major galaxies in the Local Group are surrounded by vast planes of satellites exhibiting a coherent orbital motion. The disc of satellites around M31 appears to cover $\sim 400\kpc$ in diameter, but has a thickness of $\sim 20\kpc$ and lies perpendicularly aligned with the axis joining the Milky Way and M31 (Ibata et al. 2013). In the Milky Way most satellites appear distributed over a thicker plane which is inclined by $35^\circ$ degrees to this axis (Metz et al. 2007; Pawlowski et al. 2013), although both the internal location of the Sun and the lack of deep photometric surveys of the Southern hemisphere introduce severe uncertainties in this measurement. Recently, Tully (2013) has pointed out that the presence of discs of satellites may also extend to galaxies beyond the Local Group (M81 and Centaurus A). 
Accretion of structures along dark matter filaments (Libeskind et al. 2010; Lovell et al. 2011; Shaya \& Tully 2013) and in groups (Lynden-Bell 1982; D'Onghia \& Lake 2008; Deason et al. 2011) may be plausible mechanisms to explain the planar distributions of satellites. However, filaments tend to be as extended as the virial radius of the host dark matter halo (Vera-Ciro et al. 2011), whereas disrupted associations quickly thicken unless the progenitor's orbit is perfectly aligned with the long or short axis of a triaxial halo (Bowden et al. 2013). Indeed, Bahl \& Baumgardt (2014) and Ibata et al. (2014) find that satellite planes with the thinness, radial extent and coherent kinematics as the one observed in the Andromeda galaxy are extremely rare, occurring in $\lesssim 1\%$ of the halo analogues found in the Millennium II simulations.

The apparent mismatch between $\Lambda$CDM predictions and observations has motivated the exploration of alternative gravity theories (e.g. Kroupa et al. 2010). Recently, Zhao et al. (2013) have re-derived the timing argument under the assumption that dynamics on small scales are governed by the empirical MOND law of Milgrom (1983). These authors show that as a result of the extended Mondian attraction the observed baryonic masses of the Milky Way and M31 imply a past close encounter between the two galaxies between 6 to 12 Gyr ago. Such interaction may explain the presence of extended satellite structures surrounding both galaxies (Pawlowski et al. 2013).
 
This paper approaches these issues from a $\Lambda$CDM framework. First, we inspect the Newtonian equations of motion that describe the perturbations in the Hubble flow by the Milky Way and M31 beyond the point-mass approximation (\S\ref{sec:eqmot}). Section~\ref{sec:num} presents constrained N-body experiments designed to check the assumptions on which the classical timing argument rests. 
Section~\ref{sec:like} outlines the Bayesian methodology used to analyze the observations of the local Hubble flow presented in Section~\ref{sec:obs}. Mock data are generated using the above N-body models in order to quantify the systematic behaviour of the errors identified in our method. Section~\ref{sec:results} describes the results of our analysis. A discussion of these results is given in Section~\ref{sec:discussion}. Finally, Section~\ref{sec:summary} summarizes the main findings of this contribution.

\section{The local Hubble flow} \label{sec:eqmot}
\subsection{Equations of motion}
As a result of the combined gravitational pull of the Milky Way and M31, galaxies in the vicinity of the Local Group move on orbits that deviate from the FLRW model.
At sufficiently large distances these modifications can be described as gravitational perturbations of the solutions to the Friedmann equation. To analyze the impact of the Local Group on the local Hubble flow it is useful to adopt a coordinate frame whose origin is located at the Local Group barycentre 
\begin{equation}\label{eq:com}
M_{\rm G}{\bf r}_{\rm G}+ M_{\rm A}{\bf r}_{\rm A} = {\bf 0};
\end{equation}
so that ${\bf r}_{\rm A}=-M_{\rm G}/M_{\rm A}{\bf r}_{\rm G}= -f_{\rm m} {\bf r}_{\rm G}$, where $f_{\rm m}\equiv M_{\rm G}/M_{\rm A}$ is the Milky Way to M31 mass ratio. The current separation between both galaxies is $d=|{\bf r}_{\rm A}-{\bf r}_{\rm G}|\approx 783\pm 25\kpc$ (McConnachie 2012). 

The equations of motion of a mass-less particle orbiting in the Local Group are
\begin{equation}\label{eq:eqmot}
\ddot{\bf r} =-\frac{G M_{\rm G}}{({\bf r}-{\bf r}_{\rm G})^3} ({\bf r}-{\bf r}_{\rm G}) -\frac{G M_{\rm A}}{({\bf r}-{\bf r}_{\rm A})^3} ({\bf r}-{\bf r}_{\rm A}) + H_0^2\Omega_{\Lambda}{\bf r}.
\end{equation}
At large distances ($r\gg d$), the relative separation between A and G and the tracer galaxy can be approximated as $|{\bf r}-{\bf r}_{\rm G}|\approx |{\bf r}-{\bf r}_{\rm A}|\approx r$. Thus Equation~(\ref{eq:kep}), which faithfully describes the relative motion between M31 and the Milky Way, can be also used to compute the orbits of {\it distant} galaxies with respect to the Local Group barycentre\footnote{The sample we have gathered from the literature in Section~\ref{sec:obs} includes a large number of galaxies at distances comparable to the separation between the Milky Way and Andromeda. Section~\ref{sec:num} explicitly tests the validity of the point-mass approximation with aid of N-body experiments.}. Indeed, it was originally pointed out by Lynden-Bell (1981) and Sandage (1986) that the equations on which the `timing argument' rests also govern the orbits of individual galaxies in the outskirts of the Local Group (see also Chernin et al. 2009). 

When modelling the orbits of galaxies about the Local Group barycentre we integrate the equations of motion from a time close to the Big Bang ($t_{\rm init}\approx 0$) to the current epoch $t=t_0$. The initial conditions $r(t_{\rm init})= r_\epsilon\ll d$, and $v_{\rm init}\equiv \dot r(t_{\rm init})$, are chosen to match the observed distance and radial velocity of nearby galaxies at $t=t_0$ (see \S\ref{sec:like} for further details).
 
Assuming null curvature the age of the Universe ($t_0$) follows from our choice of the cosmological parameters $H_0$ and $\Omega_\Lambda$
\begin{equation}\label{eq:age}
t_0=\int_0^{t_0}\d t=\int_0^1 \frac{\d \xi}{H \xi}= H_0^{-1}\int_0^1 \frac{\d \xi}{[(1-\Omega_\Lambda)\xi^{-3}+\Omega_{\Lambda}]^{1/2} \xi};
\end{equation}
where $H(t)=\dot \xi/\xi(t)$; $\xi=(1+z)^{-1}$ and $z$ is the redshift. Here we have used Friedmann's equation $H(\xi)=H_0[(1-\Omega_\Lambda) \xi^{-3}+\Omega_{\Lambda}]^{1/2}$ with no radiation.


For illustrative purposes we shall adopt the following fiducial parameters: $h\equiv H_0/(100 \kms\mpc^{-1})=0.7$ and $\Omega_\Lambda=0.7$. Plugging these parameters in Equation~(\ref{eq:age}) yields $t_0\simeq 13.46$ Gyr.

\subsection{Can we measure the cosmological constant locally?}\label{sec:de}
The motion of particles implied by Equation~(\ref{eq:eqmot}) can be solved analytically if one sets $\Omega_\Lambda=0$ and models the Local Group as a point mass. Under this approximation the orbits of galaxies reduce to Keplerian orbits, which can be expressed as
\begin{equation}\label{eq:rkep}
r=a(1-\cos 2 \eta);
\end{equation}
where $a=GM/(-2E)$ is the semi-major axis of the orbit, $E$ is the orbital energy, and $\eta$ is an angle typically referred to as the eccentric anomaly, which can be calculated numerically from the following equation
\begin{equation}\label{eq:eta}
2\eta -\sin 2\eta = (GM/a^3)^{1/2}t.
\end{equation}

Using Equations~(\ref{eq:rkep}) and~(\ref{eq:eta}), and defining the frequencies $\Omega=(G M/r^3)^{1/2}$ and $\omega=v/r$, Lynden-Bell (1981) showed that in a matter-dominated Universe the current locations and velocities of galaxies around the Local Group follow a close-to-linear relation
\begin{equation}\label{eq:donald}
\Omega t_0 + m \omega t_0 =n;
\end{equation}
with $m\simeq 0.9$ and $n=2^{-3/2}\pi\simeq 1.1$. Equation~(\ref{eq:donald}) is exact for $\omega=v=0$, that is $\eta=\pi/2$, which corresponds to the radius $r_0$ where the expansion of the Universe is momentarily stopped due to the attraction of the Local Group (i.e. the so-called `turn-around radius').

\begin{figure}
 \includegraphics[width=84mm]{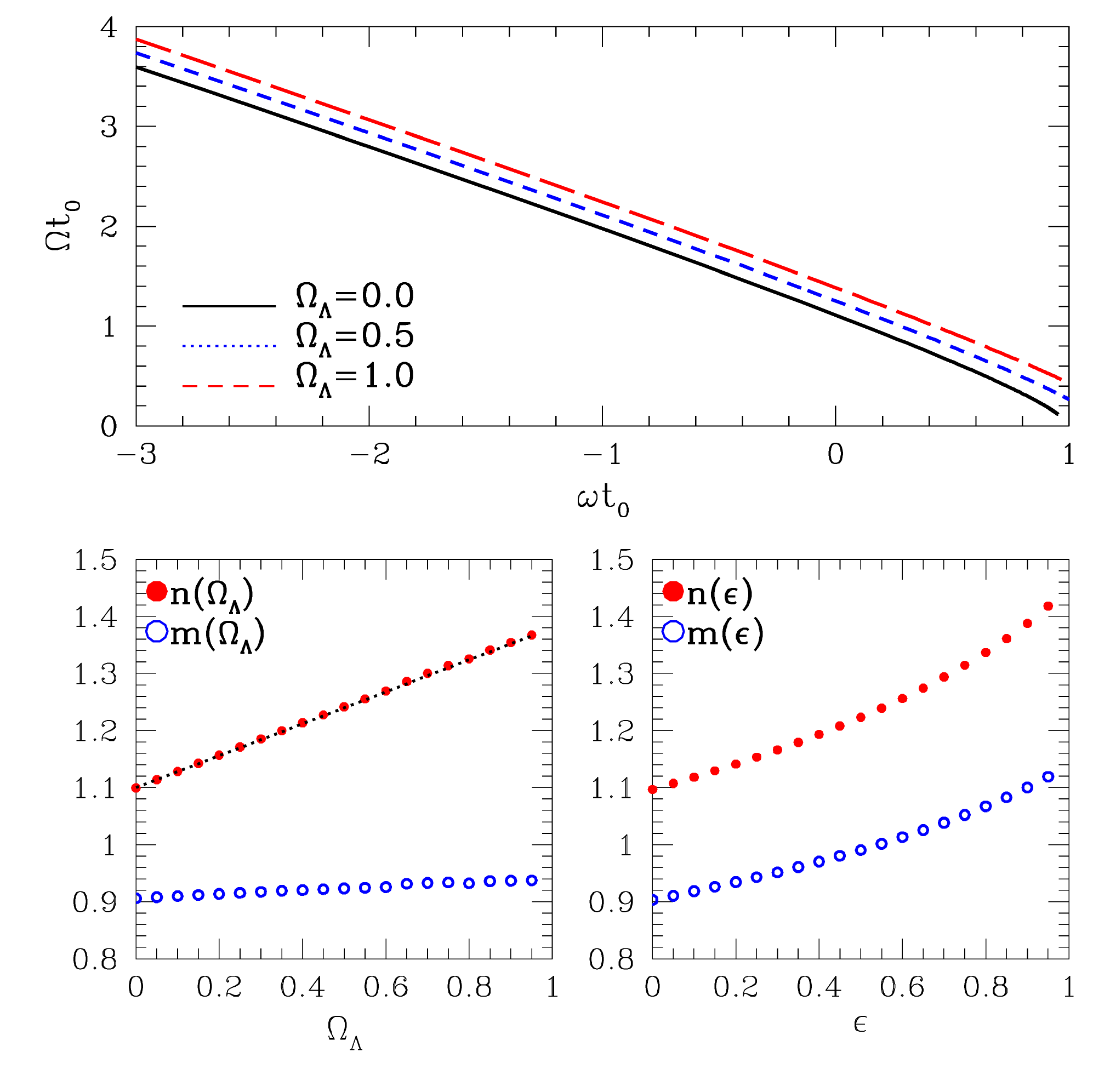}
\caption{{\bf Upper panel}: Relation between the orbital frequency $\omega=v/r$, and the frequency associated with the dynamical time, $\Omega=(G M/r^3)^{1/2}$, for different values of $\Omega_\Lambda=1-\Omega_m$. These curves are derived integrating Equation~(\ref{eq:kep}) numerically for orbits within $-3 \le \omega t_0 \le 1$. {\bf Bottom-left panel}: Linear fits to the curves shown above for a wide range of vacuum energy densities, see Equation~(\ref{eq:donald}). Note that the slope $m\simeq 0.91$ is barely sensitive to $\Omega_\Lambda$, whereas the abscissa varies as $n\simeq 1.1 + 0.28 \Omega_\Lambda$ (dotted line). 
{\bf Bottom-left right panel}: As in the left panel with a Local Group mass that varies with time as $M(t)=M_0[1+(t-t_0)\epsilon/t_0]$. For ease of comparison we set $\Omega_\Lambda=0$ in Equation~(\ref{eq:kep}).}
\label{fig:freq}
\end{figure}

Fig.~\ref{fig:freq} shows that Equation~(\ref{eq:donald}) also reproduces the relation between the position and velocity of the Local Group members in a Universe with vacuum energy. The bottom panel indicates that the slope $m$ does not depend strongly on 
the pressure term of Equation~(\ref{eq:kep}), while the abscissa $n$ slightly increases from 1.1 to 1.38 depending whether the expansion of the Universe is matter or dark-energy dominated.

Combination of Equation~(\ref{eq:donald}) with the linear relation $n=n(\Omega_\Lambda)$ fitted in the lower panel of Fig.~\ref{fig:freq} provides an analytical description of the perturbed Hubble flow in a $\Lambda$CDM Universe
\begin{equation}\label{eq:vd}
v\approx (n-\Omega t_0)\frac{r}{m t_0}\simeq 1.2 \frac{r}{t_0} - 1.1\bigg(\frac{GM}{r}\bigg)^{1/2} + 0.31 \Omega_\Lambda\frac{r}{t_0}.
\end{equation}
It is worth highlighting the small impact of the cosmological constant on the dynamics of Local Group galaxies. From Equation~(\ref{eq:vd}) the pressure-term in the equation of motion leads to a velocity increase $\delta_\Lambda\equiv v-v(\Omega_\Lambda=0)\simeq 0.31 \Omega_\Lambda r/t_0$ independently of the Local Group mass. For galaxies within $r<3 \mpc$ and $t_0=13.46$ Gyr this implies $\delta_\Lambda\lesssim 46 \kms$, which is of the same magnitude as the local velocity dispersion about the Hubble flow (see \S\ref{sec:results}). The location of the zero-velocity radius is also scarcely sensitive to the pressure term, $r_0=[GM t_0^2/n^2]^{1/3}\simeq r_0(\Omega_\Lambda=0)(1-0.17\Omega_\Lambda)$. Given that the observed value is $r_0\approx 1\mpc$ (McConnachie 2012), the cosmological constant shifts the location of $r_0$ by an amount that is comparable to the error in the distance of several galaxies in our sample (see \S\ref{sec:obs}). The small (albeit non-negligible) contribution of the cosmological constant to Equation~(\ref{eq:vd}) suggests that the shape of the local Hubble flow is mainly constrained by the combined masses of the Milky Way and M31 and the age of the Universe, as discussed below.

\subsection{Effects of a time-dependent Local Group potential}\label{sec:massev}
The analytical derivation of the timing argument given by Equation~(\ref{eq:vd}) assumes that the mass of the Local Group remains constant throughout the expansion of the local Universe. Given that this is clearly at odds with hierarchical galaxy formation theories we explore below to what extent the local Hubble flow is sensitive to time variations of the Local Group potential.

To this end we use the method of Pe\~narrubia (2013) for constructing dynamical invariants in time-dependent gravitational potentials. The technique relies on a canonical transformation ${\bf r}\mapsto {\bf r}'R$, and a time-coordinate transformation $\d t\mapsto \d \tau R^{2}$, that removes the explicit time-dependence from the equations of motion. Here ${\bf r}'(\tau)$ corresponds to orbits calculated in a static potential with $M(t)=M_0$.
For power-law forces $F(r,t)=-GM(t) r^n$, the scale factor is approximately $R(t)\approx [M_0/M(t)]^{1/(3+n)}$ (see Pe\~narrubia 2013 for details). 

In the adiabatic limit, i.e. $(\dot M/M_0)^{-1}\gg T$, where $M_0=M(t_0)$ the current Local Group mass and $T$ is the radial period of a galaxy, the evolution of the orbital energy is
\begin{eqnarray}\label{eq:ead}
E(t)= \frac{E_0}{R^2(t)} + {\cal O}\bigg(\frac{\dot R}{R}\bigg);
\end{eqnarray}
where $E_0=E(t_0)$ is the energy measured from the current position and velocity vectors of a particle. 
As expected Equation~(\ref{eq:ead}) implies no energy variation, regardless of the intermediate evolution of the system, if the final potential is the same as the initial. 

The frequencies $\Omega$ and $\omega$ appearing in Equation~(\ref{eq:donald}) can be re-written in the generic case where the Local Group evolves adiabatically
\begin{eqnarray}\label{eq:omegat}
\Omega(t)\mapsto\frac{1}{R^2}\bigg[\frac{G M_0}{r'^3}\bigg]^{1/2}=\frac{\Omega'}{R^2(t)} \\ \nonumber
\omega(t)\mapsto\frac{1}{r'}\frac{\d r'}{\d \tau}= \frac{1}{R(t)r'}\bigg[2\bigg(\frac{E_0}{R^2(t)} + \frac{GM_0}{r'R^2(t)}\bigg)\bigg]^{1/2}=\frac{\omega'}{R^2(t)};
\end{eqnarray}
where $\Omega'$ and $\omega'$ denote frequencies measured in a fixed potential with $M(t)=M_0$. Given that at $t=t_0$ the scale factor is $R(t_0)=1$, we find that in the adiabatic limit the measured Hubble flow is independent of the past evolution of the Local Group mass. 

To obtain a first-oder correction beyond the adiabatic approximation we notice that in the transformed coordinates the age of the Universe is 
\begin{equation}\label{eq:tau}
\int_0^{\tau_0} \d\tau=\int_0^{t_0}\frac{\d t }{R^2(t)}.
\end{equation}
By integrating both sides of Equation~(\ref{eq:tau}) one can define the time shift $\Delta t$ as $\tau_0 = t_0 + \Delta t$.

In the limit $\Delta t/t_0 \ll 1$ the velocity measured at a fixed radius $r(t_0)=r'(\tau_0)$ can be written as
\begin{equation}\label{eq:vdt}
v=v'(r'[\tau_0])\simeq v' (1 -\Omega\Delta t);
\end{equation}
where $v'(r')$ again corresponds to the Hubble relation if the case $M(t)=M_0$.
Equation~(\ref{eq:vdt}) indicates that the time-dependence of the gravitational force changes the radial dependence of $\omega$, while leaving $\Omega$ unperturbed. It is straightforward to show that both the slope and abscissa defined in Equation~(\ref{eq:donald}) vary by the same amount, namely $m=m'(1-m'w'\Delta t)$ and $n=n'(1-m'w'\Delta t)$.

For illustrative purposes let us consider the case where the Local Group mass varies linearly with time, $M(t)=M_0[1 + \epsilon (t-t_0)/t_0]$, with a mass growth rate $0\le \epsilon \le 1$. Recall that for a Keplerian potential $n=-2$ and the scale factor is $R(t)\approx M_0/M(t)$, so that the time shift given by Equation~(\ref{eq:tau}) is simply $\Delta t=-\epsilon t_0/2$. Given that the Local Group mass is expected to grow with time we set $\epsilon>0$, so that both the slope and abscissa increase by a factor $m'(w't_0)\epsilon/2$, as shown in the lower-right panel of Fig.~\ref{fig:freq}. 

As we shall see in Section~\ref{sec:obs}, most galaxies in our sample are moving away from the Local Group barycentre (i.e. $\omega t_0 \gtrsim 0$). According to Fig.~\ref{fig:freq} this implies $\Omega t_0 \lesssim 1$. Therefore, for a linear mass growth $\Delta v/v= \Omega t_0 \epsilon/2 \ll 1$, suggesting that local Hubble flow holds little information on the time-dependence of the Local Group potential. For simplicity the dynamical models discussed below are built in static potentials.

\begin{figure}
 \includegraphics[width=84mm]{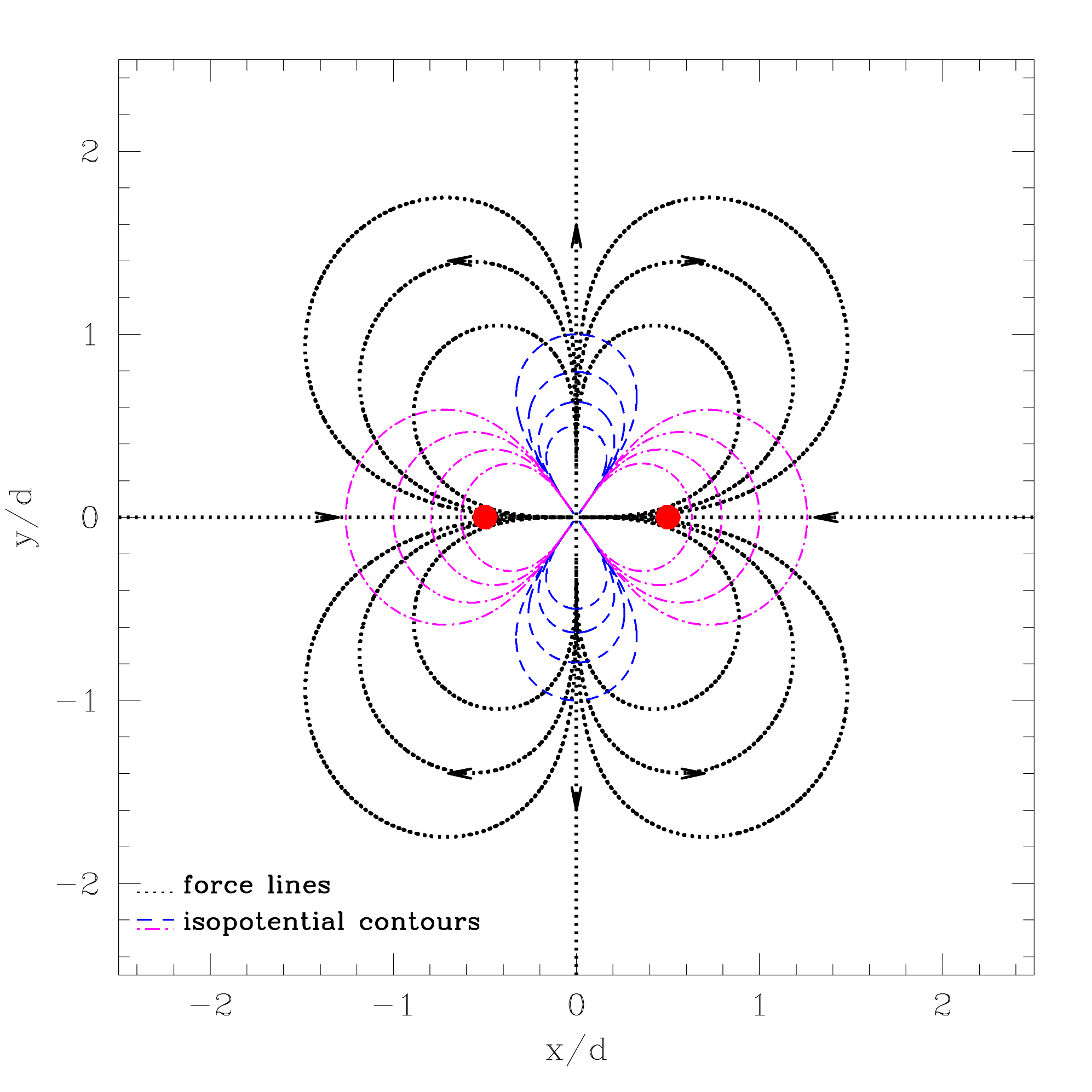}
\caption{Potential quadrupole of two equal-mass point-masses (red dots) located at ${\bf r}_{\rm A}/d=(-0.5,0,0)$ and ${\bf r}_{\rm G}/d=(+0.5,0,0)$. Iso-potential contours are colour coded according to their sign (magenta/blue for an attractive/repulsive quadrupole). Force lines are plotted with black dotted lines, with arrows marking the direction of the force (see text). }
\label{fig:quad}
\end{figure}
\subsection{The Local Group quadrupole}\label{sec:quad}
The equations of motion outlined in Section~\ref{sec:de} assume that the Local Group can be modelled as a central point mass. 
However, a far more accurate representation of the Local Group corresponds to a gravitational field dominated by the Milky Way and M31 masses. To study the gravitational potential of such a system it is useful to choose a Cartesian system where the galaxies A and G with masses $M_{\rm A}=M/(1+f_{\rm m})$ and $M_{\rm G}=M f_{\rm m}/(1+f_{\rm m})$ are located at ${\bf r}_{\rm A}=(-d/[1+f_{\rm m}],0,0)$ and ${\bf r}_{\rm G}=(+d f_{\rm m}/[1+f_{\rm m}],0,0)$. In these coordinates the gravitational potential associated with Equation~(\ref{eq:eqmot}) can be written as
\begin{eqnarray}\label{eq:pot}
\Phi(x,y,z)=-\frac{G M }{(1+f_{\rm m})\{[x+d f_{\rm m}/(1+f_{\rm m})]^2+y^2+z^2\}^{1/2}} \\ \nonumber 
-\frac{G Mf_{\rm m} }{(1+f_{\rm m})\{ [x-d /(1+f_{\rm m})]^2+y^2+z^2 \}^{1/2}}+ \frac{1}{2}H_0^2\Omega_{\Lambda}r^2.
\end{eqnarray}

Let us now consider galaxies at large distances from the Local Group barycentre, i.e. $d/r\lesssim 1$. At large radii the potential can be approximated as a Taylor expansion of~(\ref{eq:pot})
\begin{eqnarray}\label{eq:potexp}
\Phi(r,\theta)\approx -\frac{G M }{r} + \frac{1}{2}H_0^2\Omega_{\Lambda}r^2 +\frac{GM f_{\rm m}}{2(1+f_{\rm m})^2}\frac{(1-3 \cos^2\theta)d^2}{r^3};
\end{eqnarray}
where $\cos\theta\equiv x/r$. Expressing the total potential as the sum of a spherical plus axi-symmetric terms, $\Phi(r,\theta)=\Phi^{(0)}(r)+\Phi^{(2)}(r,\theta)$, we find that the right-hand term of Equation~(\ref{eq:potexp}) corresponds to a gravitational quadrupole whose strength decays as $\Phi^{(2)}\sim 1/r^3$. 

At small distances from the Local Group barycentre, $r\lesssim [4GM/(H_0^2\Omega_\Lambda)]^{1/3}$, the contribution of the quadrupole to the potential, $\Phi^{(2)}/\Phi$, does not depend on the total mass of the pair, $M$. It is, however, fairly sensitive to the mass ratio between the galaxies A and G. For example, the quadrupole term 
reaches its maximum if the galaxies have equal masses (i.e. $f_{\rm m}=1$), and 
approaches zero asymptotically if one of the galaxy pair dominates the total mass (i.e. either $f_{\rm m}\rightarrow 0$, or $f_{\rm m}\rightarrow \infty$). 

Dashed lines in Fig.~\ref{fig:quad} follow the iso-potential contours of the quadrupole
\begin{eqnarray}\label{eq:quad}
\Phi^{(2)}=\Phi^{(2)}_0 \frac{(1-3 \cos^2\theta)d^3}{r^3},
\end{eqnarray}
where $\Phi^{(2)}_0 =GM f_{\rm m}/[2d(1+f_{\rm m})^2]$. Contours in this plot correspond to $\rho=1, 2, 4$ and 8 in the equation $r(\rho,\theta)=d|1-3 \cos^2\theta|^{1/3}\rho^{1/3}$. Notice that the sign of $\Phi^{(2)}$ flips at $\theta=\cos^{-1}(1/\sqrt{3})\approx 54.73^\circ$. For ease of reference contours are colour-coded according to the quadrupole sign (blue/magenta for positive/negative values). 

Fig.~\ref{fig:quad} also shows Faraday's {\it lines of force} (black dotted lines). These lines provide a useful representation of the quadrupole, as the number of lines at a given point is related to the strength of the field, whereas the tangent of any curve at a particular point is oriented along the direction of the force (marked with arrows for reference). Each line of force corresponds to a solution to the differential equation $r\d\theta/\d r=F_\theta/F_r$, where $F_r=\partial \Phi^{(2)}/\partial r$ and $F_\theta=r^{-1}\partial \Phi^{(2)}/\partial \theta$. From Equation~(\ref{eq:quad}) we find that the line of forces follow the equation $r/d=\rho'\cos^{1/4}\theta\sin^{1/2}\theta$. 

Note that the gravitational quadrupole induces an attractive force along the axis joining galaxies A and G which becomes repulsive in any transverse direction. Given the azimuthal symmetry of the field it is convenient to write the force in polar coordinates aligned with this axis
\begin{eqnarray}\label{eq:forceq}
F_\parallel=F_x=-\frac{\partial \Phi^{(2)}}{\partial x}= \frac{3\Phi^{(2)}_0d^3}{r^4}(2\cos^2\theta-3\sin^2\theta)\cos\theta\\ \nonumber
F_\perp=\sqrt{F_y^2+F_z^2}=\frac{3\Phi^{(2)}_0d^3}{r^4}(4\cos^2\theta-\sin^2\theta)\sin\theta.
\end{eqnarray}
The force along the symmetry axis is twice as large as in any perpendicular direction and has an opposite sign, i.e. $|F_\parallel(\theta=0)|=-2|F_\perp(\theta=\pi/2)|$. 

These results suggest that the Local Group quadrupole may have a sustained impact on the orbits of galaxies that define the {\it local} Hubble flow, a possibility that we inspect below with the aid of N-body experiments.

\begin{figure*}
\vspace*{0cm}\includegraphics[width=180mm]{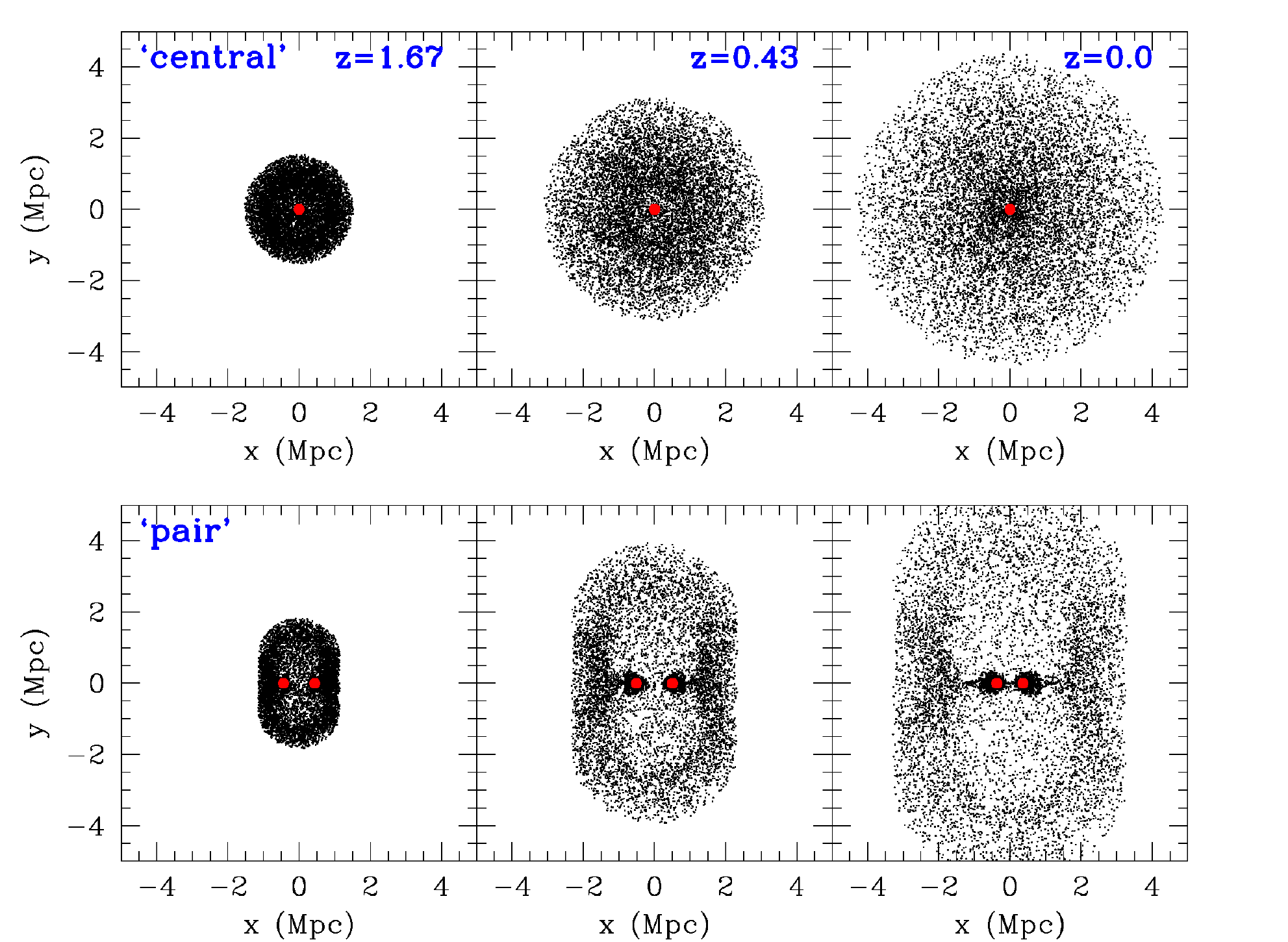}
\caption{Snap-shots of the distribution of particles along an axis perpendicular to the relative motion between the galaxies A and G (red dots). The upper panels adopt a Local Group model where all the mass is located at the barycentre. In the lower panels the Local Group mass is made up by the galaxies A and G, which move on trajectories defined by Equation~(\ref{eq:kep}). In both cases we adopt $M=2M_{\rm A}=2M_{\rm G}=5\times 10^{12}{\rm M}_\odot$, $h=0.7$ and $\Omega_\Lambda=0.7$. 
The separation between A and G at $t=t_0$ is $d\approx 0.78\mpc$. In both experiments the initial velocity distribution of the tracer particles (black dots) is isotropic. Yet, in models where the potential is dominated by a galaxy pair the infall of particles occurs preferentially along the axis joining both galaxies. Comparison with Fig.~\ref{fig:quad} shows that the distribution of particles traces the lines of force associated with the potential quadrupole.}
\label{fig:rr}
\end{figure*}

\begin{figure*}
\vspace*{0cm}\includegraphics[width=180mm]{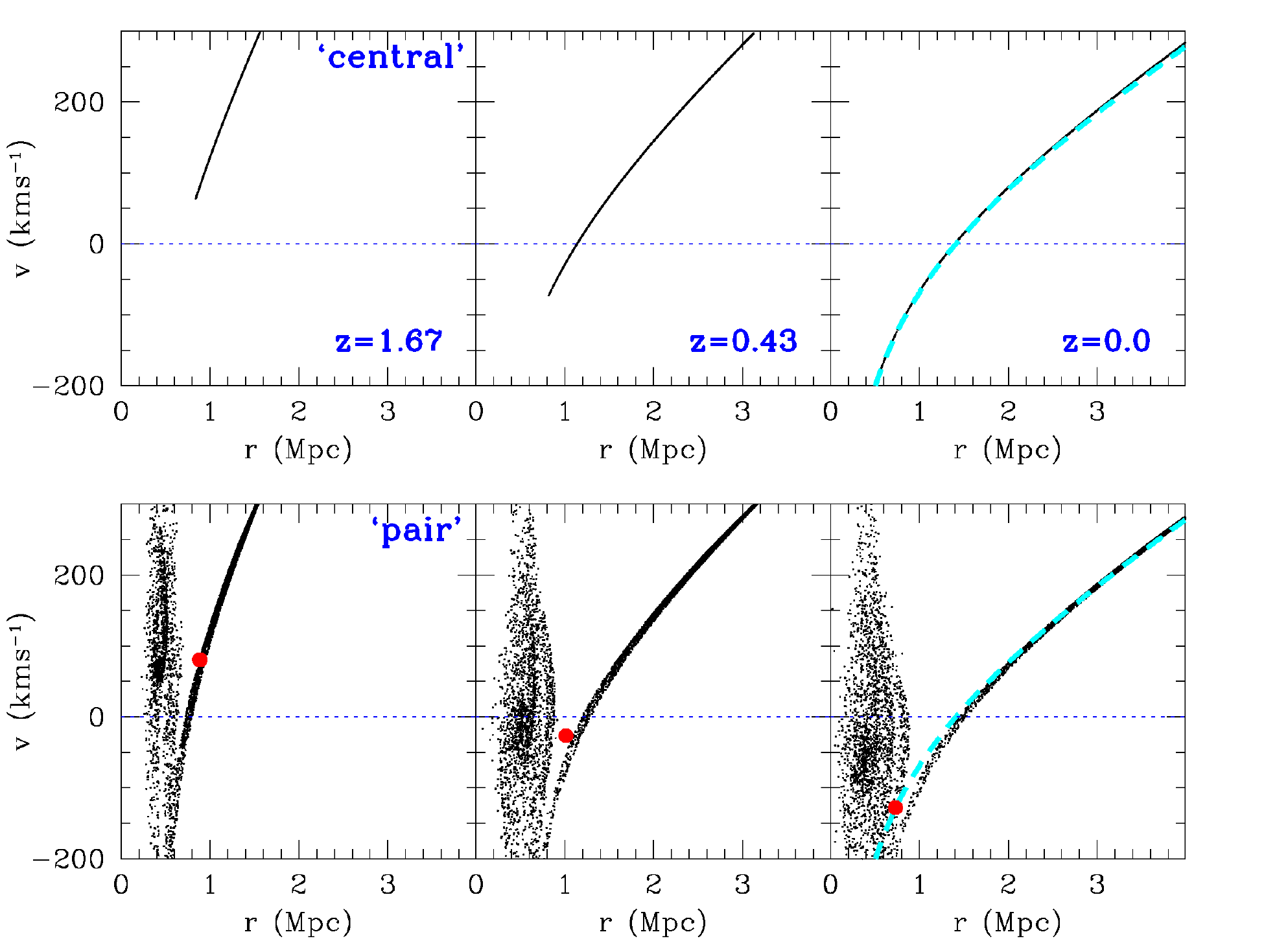}
\caption{Hubble flows associated with the models shown in Fig.~\ref{fig:rr}. Red dots mark the relative distance and velocity between the galaxies A and G at each snap-shot. Their separation at $z=0$ is $d(t_0)=0.78\mpc$. By construction the Hubble flow of the `central' Local Group model (upper panel) is an exact solution to Equation~(\ref{eq:kep}) (cyan dashed curves). In contrast the lower panels show that the potential quadrupole causes strong perturbations on the kinematics of nearby galaxies. Some of the particles that fall back toward the Local Group barycentre become bound to the galaxies A and G, which leads to a large velocity scatter at $r\lesssim d(t_0)$. Note also that at intermediate distances ($1 \lesssim r/d(t_0)\lesssim 3$) the stronger gravitational pull along the axis joining A and G tends to increase the infall velocity of particles with respect to solutions to Equation~(\ref{eq:kep}). }
\label{fig:vv}
\end{figure*}

\section{Restricted N-body models}\label{sec:num}
In this Section we carry a suite of N-body experiments that follow the expansion of the local ($<4\mpc$) Universe from a time close to the Big Bang to the present. Although these experiments do not capture the complexity of the non-linear growth of structures in a $\Lambda$CDM cosmology, they do share essential features with the hierarchical formation of galaxies in an expanding Universe and provide useful insight into the perturbations in the Hubble flow by the Local Group.

The initial conditions are set up so that the final configuration of particles can be approximately described by a FLRW model, where all bodies move away from each other on radial orbits isotropically distributed in space. To this end we place all particles initially at a radius $r=r_\epsilon$. Subsequently, orbital energies are randomly generated within the interval $E_{\rm min}< E_i < E_{\rm max}$, with the range chosen so that  at $t=t_0$ test particles are homogeneously distributed within $0\lesssim r/\mpc \lesssim 4$. For each particle the velocity associated with $E_i$ is $v_{{\rm init},i}=[2(E_i+GM/r_\epsilon)]^{1/2}$. The directions of the velocity vectors are randomly distributed on the surface of a sphere. 

Each combination of $(r_\epsilon,v_i)$ defines an orbit which is integrated from $t_{{\rm init},i}=r_\epsilon/v_{{\rm init},i}$ to the present, $t=t_0$. This is done through a leap-frog integration of Equation~(\ref{eq:kep}), with a time-step chosen so that energy is conserved at a $10^{-3}$ accuracy level. Due to the central divergence of the Keplerian potential the value of $r_\epsilon$ cannot be arbitrarily close to zero. Yet, the choice of $r_\epsilon$ should not influence the properties of the Hubble flow at $t=t_0$. We find that $r_\epsilon=0.04 \mpc$ is sufficiently small so that this condition is met with ease.

Upper panels in Fig.~\ref{fig:rr} show three snap-shots of the expansion of an idealized (local) Universe. Small black dots correspond to mass-less (`dust') particles that move on a Keplerian ``central'' potential with $M=5\times 10^{12}{\rm M}_\odot$ (thick red dot). At early stages all particles move with very high velocities and occupy a densely-packed volume. As the Universe expands the mean density of particles decreases in a monotonic fashion. The gravitational pull of the Local Group slows down the motion the particles nearest to the central galaxy, so that eventually a fraction of them reach a turn-around radius and start to fall back onto the central regions of the potential. As expected, our initial conditions lead to a local universe at $t=t_0$ that resembles a quasi-homogeneous sphere. 
 
The lower panels of Fig.~\ref{fig:rr} show that a bipolar mass distribution in the Local Group breaks the underlying symmetry built in the initial conditions. 
To construct this experiment we replace the central point-mass by a ``pair'' of point-masses\footnote{Our force calculation includes a softening-length $\epsilon=5\kpc$ to avoid a divergence of the force during close encounters. We have checked that the choice of $\epsilon$ does not lead to qualitative changes in the results.}.
The relative distance between the galaxies A and G evolves according to Equation~(\ref{eq:kep}). The initial separation corresponds to the initial distance of the tracer particles, i.e. $d(t_{{\rm init}})=r_\epsilon$, and the relative velocity $v_{\rm init}$ has been chosen so that the final separation is $d(t_0)=0.78\mpc$. 

Comparison with Fig.~\ref{fig:quad} shows that 'dust' particles behave in a manner akin to the alignment of iron fillings with a magnetic field, i.e. they distribute along the lines of force defined by the potential quadrupole. Hence, we find tracer particles preferentially along the axis that joins A and G, which is the direction where the gravitational quadrupole of the Local Group is strongest. In contrast, the density of particles drops in transverse directions to the axis defined by the main galaxies, where the quadrupole force has a positive (repulsive) sign. Notice that some of the bodies that fall back toward the Local Group barycentre become bound to either A or G, inducing a strong anisotropy in the spatial distribution of particles around the two main galaxies. 

Fig.~\ref{fig:vv} shows the Hubble flows associated with the snap-shots plotted in Fig.~\ref{fig:rr}. By construction the phase-space distribution of particles in a `central' model (upper panels) is an exact solution to Equation~(\ref{eq:kep}) (cyan dashed curves). Equation~(\ref{eq:vd}) can be used to determine the radius at which the Universe expansion halts owing to the gravitational pull of the Local Group, i.e. $r_0\approx (0.7 GM t_0^2)^{1/3}\simeq 1.42\mpc$.
In comparison the lower panels exhibit a remarkable contrast. In these models the Local Group mass is made up by the combined masses of the galaxy pair, whose relative distance and velocity is marked with red dots for ease of reference. We find that the bipolar mass distribution induces strong perturbations in the local flow. In particular, the scattered velocity distribution found at small radii, $r \lesssim d(t_0)$, results from particles that become bound to either of these galaxies as they fall back toward the Local Group barycentre. In spite of the visible impact of the potential quadrupole on the distribution of tracer particles, comparison with the cyan dashed line shows that Equation~(\ref{eq:kep}) still provides a reasonable match to the kinematics of galaxies at $r \gtrsim d(t_0)$. It is worth bearing in mind, however, that this equation systematically underestimates the radial velocity of galaxies at intermediate distances, $d(t_0) \lesssim r \lesssim  r_0$. This results from the strong spatial anisotropy of tests particles, which tend to be found along the bipolar direction where the gravitational attraction is enhanced.

\section{Bayesian analysis}\label{sec:like}
In this Section we describe the fundamentals of our Bayesian analysis of the local Hubble flow and perform a number of tests using the N-body experiments outlined in Section~\ref{sec:num}. Our models contain six free parameters that are fitted simultaneously to the data. These are the Local Group mass, $M=M_{\rm G}+M_{\rm A}$, the mass ratio between the Milky Way and Andromeda, $f_{\rm m}=M_{\rm G}/M_{\rm A}$, the circular velocity of the Milky Way at the solar radius, $V_0=V_c(R_\odot)$, the reduced Hubble constant, $h$, and the fractional vacuum energy density, $\Omega_\Lambda$.

Our choice of $h$ and $\Omega_\Lambda$ as independent quantities os motivated by recent papers which show that the value of the Hubble constant is sensitive to environment (Aragon-Calvo et al. 2011; Wojtak et al. 2013), whereas the age of the Universe and the cosmological constant are not. 
This approach deviates from the standard `timing argument' described in \S\ref{sec:eqmot}, which typically assumes that the age of the Universe is a known quantity. In theory, whether we consider $\Omega_\Lambda$ and $h$, or $\Omega_\Lambda h^2$ and $t_0$, as free parameters is a subjective decision which should not have a measurable impact on our fits.  
In practice, the choice of priors can in some cases modify the posterior distributions. Following the suggestion of the anonymous referee we have explicitly checked that the bounds derived in \S\ref{sec:results} and summarized in Table~\ref{tab:results} are independent of the combination of free cosmological parameters.

\subsection{Likelihood function}\label{sec:likeform}
Consider the Gaussian likelihood function 
\begin{eqnarray}\label{eq:like}
\mathcal{L}(\{D_i,l_i,b_i,V_{h,i}\}^{N_{\rm sample}}_{i=1}|\vec{S})=\\ \nonumber
\prod_{i=1}^{N_{\rm sample}}\frac{1}{\sqrt{2\pi \sigma_i^2}}\exp\bigg[-\frac{(V_i-V_{h,i})^2}{2\sigma_i^2}\bigg];
\end{eqnarray}
where $\vec{S}=(M,f_{\rm m},V_0,h,\Omega_\Lambda,\sigma_m)$ is a vector that comprises the model parameters; $D$ and $V_h$ are heliocentric distances and velocities, respectively, and $(l,b)$ the Galactocentric coordinates.

For a given set of parameters, $\vec{S}$, our model returns a heliocentric velocity $V$ at the location $(D,l,b)$. 
The velocity of the $i$-th galaxy, $V_i$, is calculated according the following procedure: (i) first, the distance of a galaxy to the Local Group barycentre, $r_i(t_0)$ is derived using the coordinate transformation of Appendix A. Note that this conversion is model dependent, as it requires setting the values of $f_{\rm m}$ and $V_0$. (ii) Subsequently, we choose an initial (small) radius $r_\epsilon$ and solve for the initial velocity $v_{\rm init}(t=t_{\rm init})$, with $t_{{\rm init}}=r_\epsilon/v_{\rm init}\ll t_0$, so that $r(t_0$) matches the value obtained in step (i). This is done through a leap-frog integration of Equation~(\ref{eq:kep}) from $t=t_{{\rm init}}$ until $t=t_0$, with a time-step chosen so that energy is conserved at a $10^{-3}$ accuracy level (see \S\ref{sec:num}). The model parameters that enter in this step are $M, h$ and $\Omega_\Lambda$. (iii) To derive the value of $V_i$ that goes in the likelihood function we convert the Local Group-centric coordinates of the galaxy into heliocentric ones using the transformation outlined in Appendix A.

\begin{table}
\centering
\renewcommand{\tabcolsep}{0.3cm}
\renewcommand{\arraystretch}{1.5}
\begin{tabular}{| l c c l|}
\hline
\hline
Parameter & Min.  & Max. & Description \\
\hline
$M/(10^{12}{\rm M}_\odot)$       & 0.1      & 20.0    & $M=M_{\rm G}+M_{\rm A}$ \\
$\log_{10}f_{\rm m}$                     & -1.0      & 1.0    & $f_{\rm m}=M_{\rm G}/M_{\rm A}$ \\
$V_0/\kms$                & 100    & 400   & MW's circular velocity at $R_\odot$ \\
$\Omega_\Lambda$           & 0.0      & 1.0     & Fractional vacuum density \\
$h$                       & 0.2      & 1.0     & Reduced Hubble constant \\
$\sigma_m/\kms$           & 1.0      & 200    & Hyperpar. (no phys. meaning) \\
\hline
\end{tabular}
\caption[]{Boundaries of uniform priors for free parameters in our likelihood function.}
\label{tab:priors}
\end{table}

\subsection{Peculiar motions}
Equation~(\ref{eq:kep}) is built upon the assumption that all galaxies in the Local Group move on radial orbits. Deviations from this assumption contribute to the presence of a {\it peculiar} motion with respect to the Hubble flow.

It is useful to outline the main processes that may contribute to peculiar motions. First, one should consider statistical fluctuations in the data, such as the presence of observational errors and systematic biases, which typically arise from the complexities involved in the measurement of distances and velocities for such distant objects, as recently shown by Fraternali et al. (2009). Second, the presence of galaxy associations in our tracer population, which tend to be relatively common in the local volume (e.g. Tully et al. 2006; Bellazzini et al. 2013; Fattahi et al. 2013; Chapman et al. 2013), is a non-negligible source of peculiar motions. Galaxies belonging to an association move around a common barycentre, which tends to inflate the distance-velocity relation derived purely from the cosmological expansion. Also, nearby galaxy groups may perturb the motion of kinematic tracers in the periphery of the Local Group.

Unfortunately, it is difficult to determine on observational grounds which objects have been acted on by external tidal fields or by encounters with neighbour systems. Given this strong limitation, here we follow a {\it statistical} approach in order to incorporate peculiar motions in our fits. Our method relies on the assumption that the above perturbations are small and do not lead to systematic biases in the distance-velocity relation.

To account for the effect of peculiar motions we incorporate the hyperparameter $\sigma_m$ in our analysis. Its squared value is added linearly to the observational variance in order to minimize co-variance with the rest of model parameters. Hyperparameters provide a useful tool to assign weights to data sets beyond those derived from statistical errors (e.g. Hobson et al. 2002). Indeed, by introducing $\sigma_m$ in our likelihood function we allow that the observed scatter in the distance-velocity relation may not be fully accounted by random errors in the data set. The role of $\sigma_m$ is thus analogous to a `nuisance' parameter. As such, we marginalize over $\sigma_m$ in order to obtain a joint estimation of the parameters of interest.  

The two-dimensional variance of the $i$-th measurement is calculated as follows 
\begin{equation}\label{eq:var}
\sigma_i^2=\epsilon_{V,i}^2 + \epsilon_{D,i}^2 \bigg(\frac{\d V}{\d D}\bigg)^2_{(D_i,l_i,b_i)} + \sigma_m^2;
\end{equation}
 which includes velocity ($\epsilon_{V}$) as well as distance ($\epsilon_{D}$) errors (see Ma et al. 2013), plus the additional `freedom' provided by the hyperparameter $\sigma_m$.  Here $\d V/\d D$ denotes the gradient of the Hubble flow at a given location $(D,l,b)$ returned by our model.

We apply a nested-sampling technique (Skilling 2004) in order to calculate posterior distributions for our parameters and the evidence of the model. In particular we use the code {\sc MultiNest}, a Bayesian inference tool which also produces posterior samplings and returns error estimates of the evidence (see Feroz \& Hobson 2008, 2009 for details). 
Unless otherwise indicated, all our measurements adopt flat priors over ranges that include reasonable parameter values (see Table~\ref{tab:priors}).

\begin{figure}
 \includegraphics[width=84mm]{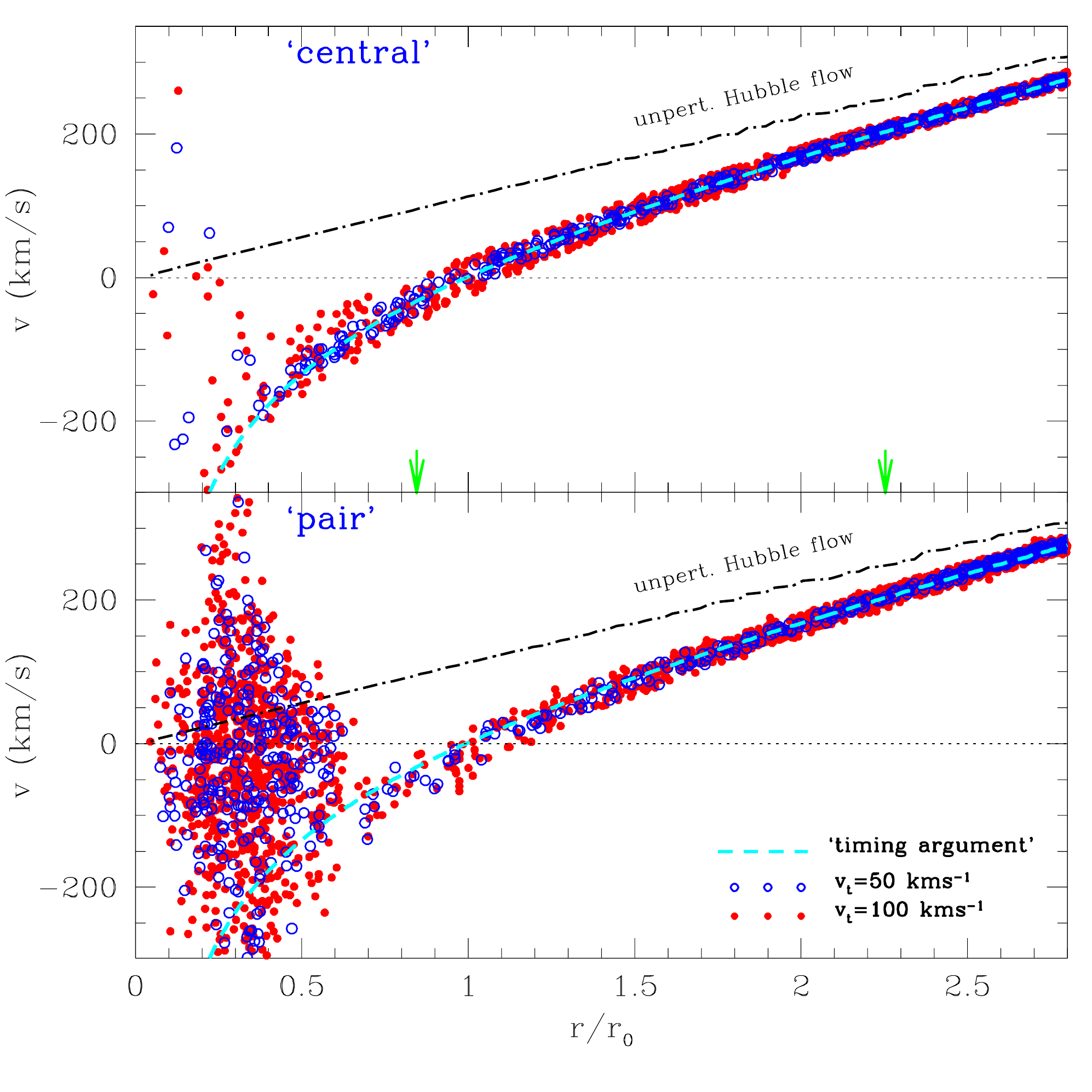}
\caption{Mock Hubble diagrams generated from the models shown in Figs.~\ref{fig:rr} and~\ref{fig:vv}. Distances are given in units of the zero-velocity radius $r_0\approx 1.42\mpc$. Black dotted-dashed and cyan dashed lines show isochrones derived from Equation~(\ref{eq:kep}) for $M=0$ (unperturbed Hubble flow) and $M=5\times 10^{12}{\rm M}_\odot$, respectively, in a Universe with $\Omega_\Lambda=0.7$ and $h=0.7$. Dots correspond to {\it estimates } of the radial velocity in a Local Group-centric frame (see Appendix A) after introducing Gaussian errors in heliocentric distance ($\epsilon_D=50\kpc$), velocity ($\epsilon_v=5\kms$), as well as a randomly-oriented tangential velocity component with moduli $v_{\rm t}=50$ (blue open dots) and $100\kms$ (red filled dots). Note that the presence of peculiar velocities plus observational errors tends to increase the scatter in the observed Hubble flow. The derivation of radial velocities becomes progressively less accurate as the distance to the barycentre decreases. The large velocity scatter in the lower panel arises from particles bound to the main galaxies. Mock data is constructed with particles within the distance range marked with green arrows.}
\label{fig:mock}
\end{figure}

\subsection{Mock data}
Prior to applying the technique outlined in the previous Section to actual data, we must examine the reliability of our method when it operates on realistic data sets that violate our model assumptions. With this aim in mind we use the models outlined in \S\ref{sec:num} to construct synthetic data sets which shall help us us to detect possible biases in the statistical inference of the model parameters. Below we describe how mock data sets are generated and the systematic behaviour of the errors identified in our method.

\subsubsection{The effect of peculiar velocities}\label{sec:pec}
We consider two sources of peculiar velocities in our tests: those arising from observational errors, which are assumed to be Gaussian, plus a transverse component, $v_t=\sqrt{v_\theta^2 + v_\phi^2}$, where $\phi$ and $\theta$ are respectively the azimuthal and polar angles in spherical coordinates, which we add to the orbital velocity vector at a random direction. We explore models with transverse velocity components comparable to the radial one, $v_t=50\kms$ and $100\kms$. Subsequently, from the position and velocity vectors in a Local Group-centric frame we derive heliocentric distances and velocities (see Appendix A), which are then convolved with Gaussian errors. In particular, we set $\epsilon_D=50\kpc$ and $\epsilon_v=5\kms$, which are broadly consistent with the observational errors in the sample of galaxies gathered from the literature (see \S\ref{sec:obs}). 

Fig.~\ref{fig:mock} shows that the main effect of peculiar velocities is to thicken the distance-velocity relation derived from the coordinate transformation of Appendix A, which rests upon the assumption that all galaxies in and around the Local Group move on radial orbits. This is because the contribution of peculiar motions to the heliocentric velocity corresponds to the projection of the transverse component onto the line-of-sight vector. The fact that our galaxies are distributed over a very large area of the sky leads to a large range of projection angles, and thus to a `scattered' contribution of peculiar motions to the bulk flow. Owing to a well-known geometrical effect the apparent scatter wanes progressively as the distance to the barycentre increases.

\begin{figure}
\vspace*{0cm}\includegraphics[width=84mm]{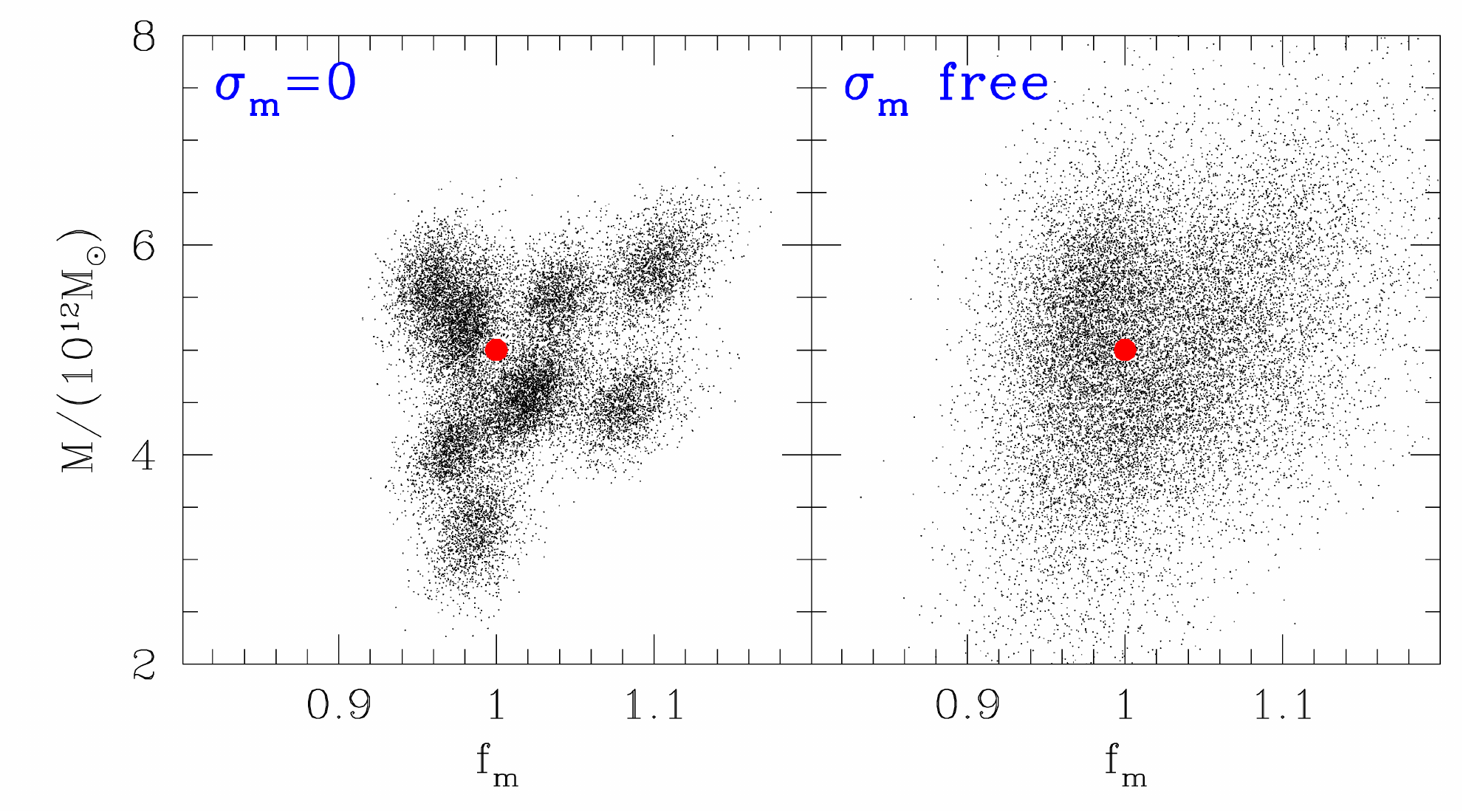}
\caption{Sampling of the posterior distributions of $M$ and $f_{\rm m}$ calculated for 10 mock data sets built from the 'central' model, each containing 35 galaxies located within a radial range $0.85\le r/r_0\le 2.25$ (see Fig.~\ref{fig:mock}). We consider errors in the heliocentric distances ($\epsilon_D=50\kpc$) and velocities ($\epsilon_v=5\kms$), plus a randomly-oriented tangential velocity component with a magnitude $v_{\rm t}=100\kms$. Red dots mark true parameter values. Comparison between the left and right panels illustrate the effects of introducing the hyperparameter $\sigma_m$ in the variance. Allowing the presence of a dispersion in our models yields more realistic uncertainties in the fitted parameters.}
\label{fig:cloud_m_fm}
\end{figure}

Fig.~\ref{fig:cloud_m_fm} illustrates how the hyperparameter $\sigma_m$ helps to lessen the impact of peculiar motions on our analysis. Here we plot posterior samplings for 10 independent mock data sets, each comprising 35 galaxies within a radial range $1.2\le r/\mpc\le 3.2$ and a transverse velocity $v_{\rm t}=100\kms$. In units of the zero-velocity radius the radial range corresponds to $0.85\le r/r_0\le 2.25$. The left panel shows that setting the parameter $\sigma_m=0$ leads to uncertainties in $M$ and $f_{\rm m}$ that are unrealistically small, as many of the probability clouds appear isolated and do not overlap with the true values. 
In contrast, fitting $\sigma_m$ and marginalizing over it yields uncertainties that do comprise the true values and can be therefore deemed more realistic.

\begin{figure}
 \includegraphics[width=84mm]{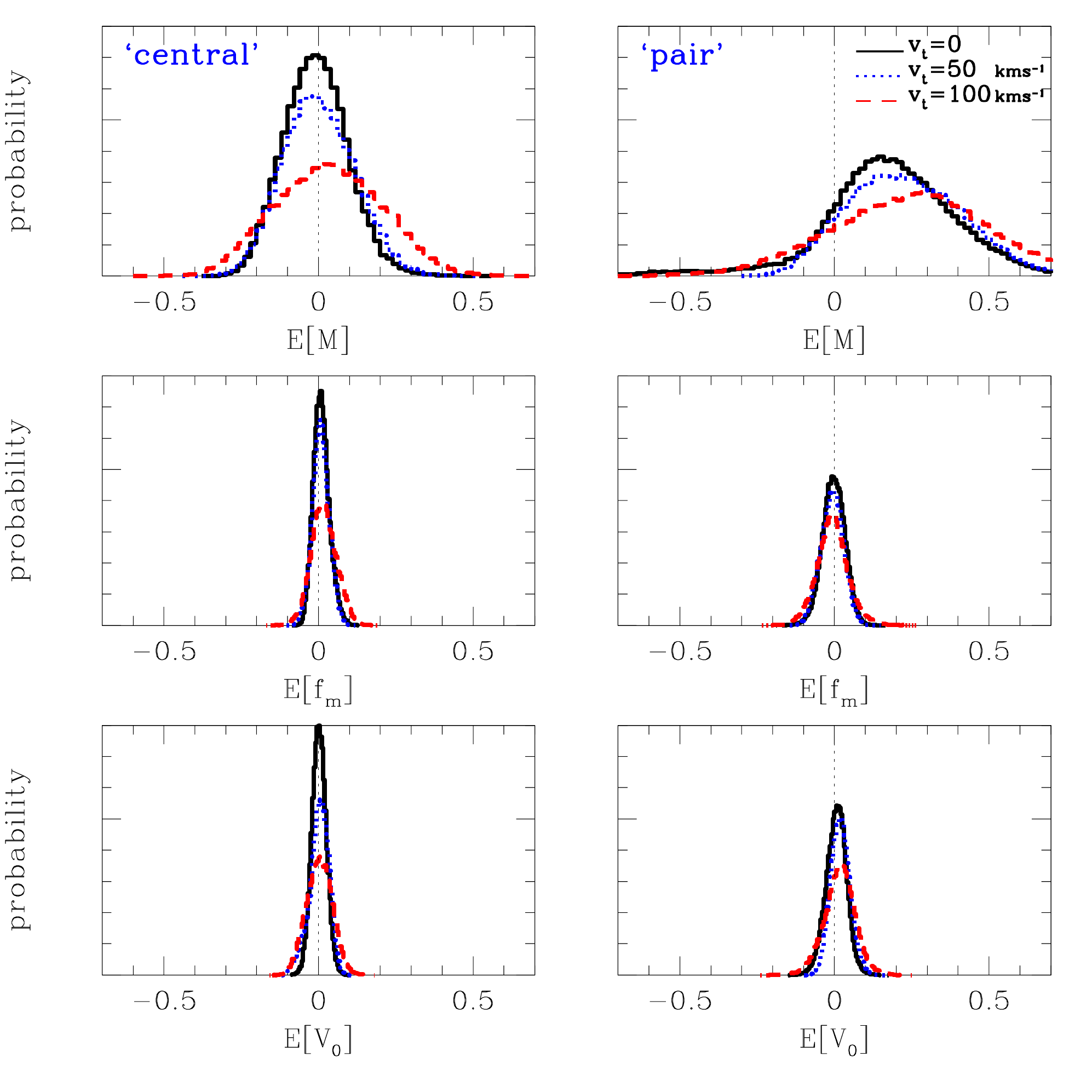}
\caption{Errors in the inference of the model parameters $M$, $f_{\rm m}$ and $V_0$, where $E(x)\equiv (x-x_{\rm true})/x_{\rm true}$ and $M_{\rm true}=5\times 10^{12}{\rm M}_\odot$, $f_{m,{\rm true}}=1$ and $V_{0,{\rm true}}=220\kms$ derived from fits of 50 independent sets of 35 mock galaxies within the distance interval $0.85\le r/r_0\le 2.25$ (see \ref{fig:mock}). Note that although peculiar velocities tend to increase the uncertainty of the fits, they do not lead to significant biases in the inferred parameters. In contrast, neglecting the contribution of the potential quadrupole in Equation~(\ref{eq:kep}) tends to over-estimate the Local Group mass by $20$--$30\%$.  }
\label{fig:err}
\end{figure}

\begin{figure}
\vspace*{0cm}\includegraphics[width=84mm]{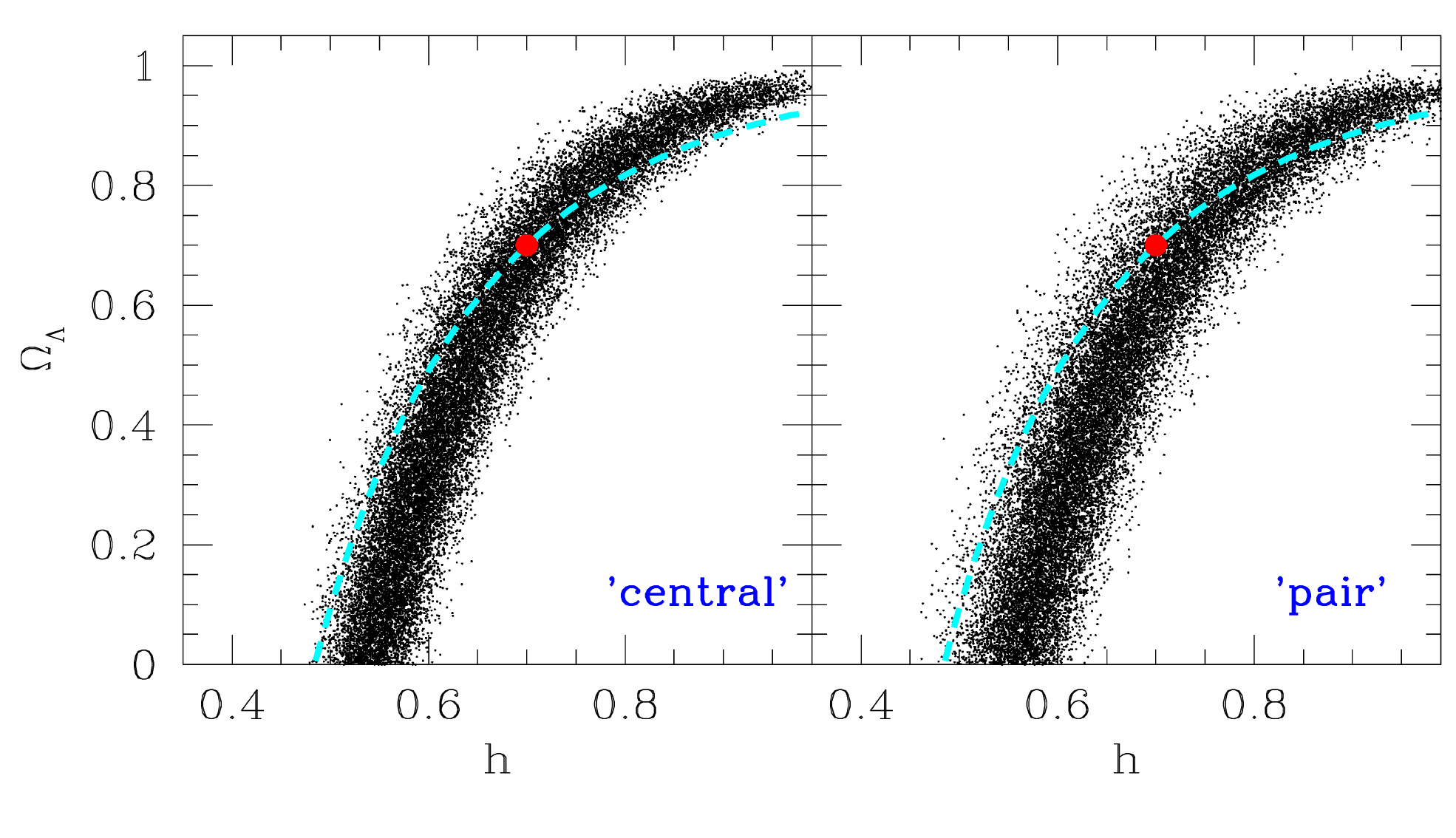}
\caption{Sampling of the posterior distributions of $\Omega_\Lambda$ and $h$ derived from 50 mock data sets with with $v_{\rm t}=100\kms$ (see Fig.~\ref{fig:err}). Red dots mark the true parameter values $h=\Omega_\Lambda=0.7$. Note that the strong degeneracy between $\Omega_\Lambda$ and $h$ roughly traces the isochrone curve $t_0=t_0(h,\Omega_\Lambda)=13.46$ Gyr derived in a Universe with null curvature (cyan dashed line). }
\label{fig:cloud_ol_h}
\end{figure}

\subsubsection{Milky Way-M31 quadrupole}\label{sec:quadeff}
Thus far we have assumed that the equations that define the motion of nearby galaxies around the Local Group are those on which the timing argument rests, i.e. Equation~(\ref{eq:kep}). Recall, however, that this approximation is only accurate for distant galaxies, i.e. in the limit $r\gg d\equiv |{\bf r}_{\rm A}-{\bf r}_{\rm G}|$. Unfortunately, we shall see in \S\ref{sec:obs} that the majority of galaxies in the galaxy sample do not obey this condition, as their distances from the Local Group barycentre tend to be comparable to the current separation between the Milky Way and Andromeda. Following the results of the restricted N-body experiments outlined in \S\ref{sec:num}, which suggest that a bipolar mass distribution in the Local Group may introduce visible effects on the kinematics of nearby galaxies, we explicitly explore here whether the point-mass approximation may be a source of systematic biases in our analysis.

The lower panel of Fig.~\ref{fig:mock} shows mock Hubble diagrams associated with the `pair' models for different values of $v_t$. 
The first point to notice is the large velocity scatter at $r\lesssim d(t_0)$, which results from the particle population bound to either of the two main galaxies. Second, we also find that the accelerated infall velocity of accreting particles along the axis joining the main galaxies leads to systematic deviations from the `timing-argument' relation (cyan dashed line). In particular, Equation~(\ref{eq:kep}) tends to under-estimate the magnitude of infall velocities at intermediate distances $d(t_0)\lesssim r\lesssim r_0$, where $r_0\approx 1.42\mpc$ is the zero-velocity radius.

Fig.~\ref{fig:err} shows the error distributions, $E(x)=(x-x_{\rm true})/x_{\rm true}$, obtained from fitting 50 mock data sets generated from the models shown in Fig.~\ref{fig:mock}. Each data set contains 35 galaxies within a distance range $0.85\le r/r_0\le 2.25$.
Focusing first on the 'central' model, we find that increasing the magnitude of the tangential velocity component tends to augment the uncertainty in $M$, $f_{\rm m}$ and $V_0$, without leading to any visible bias. In contrast, our analaysis returns slightly over-estimated masses when applied to the `pair' models. This bias increases with the magnitude of the peculiar velocity component. For the range of transverse motions explored here, we find that $M$ can be over-estimated by $20$--$30\%$. 
Remarkably, the parameters $f_{\rm m}$ and $V_0$ are measured with high precision in both sets of models. Such tight constraints result from the strong sensitivity of the solar apex to the parameter $f_{\rm m}$ and $V_0$, as discussed in Appendix B.

\begin{figure*}
\vspace*{0cm}\includegraphics[width=180mm]{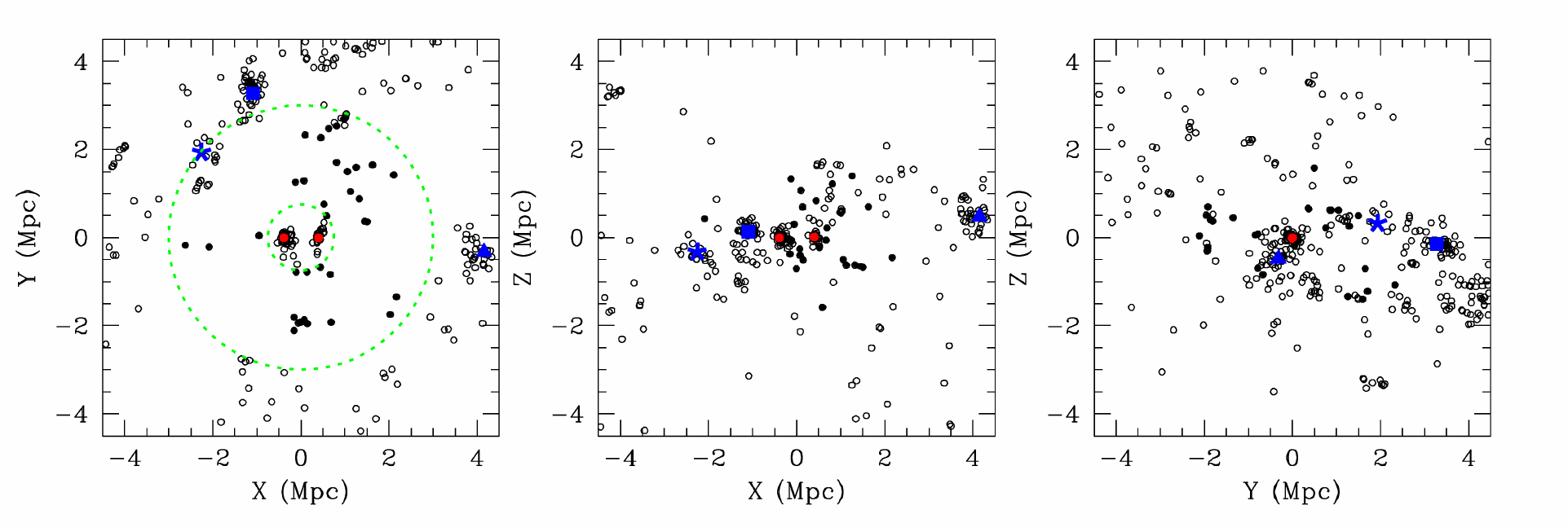}
\caption{Spatial distribution of galaxies within $5\mpc$. The Milky Way and M31 are marked with red dots. We use Cartesian coordinates where the axis joining the Milky Way and M31 is aligned with the X-axis. Dashed lines in the left panel mark $0.75$ and $3.0\mpc$ radii from the Local Group barycentre, which is derived under the assumption that both galaxies have equal masses. For ease of reference, we highlight the main associations/clusters around the Local Group, namely IC 342 (blue asterisk), M81 (blue square) and Centaurus A (blue triangle). Fill dots denote galaxies within the dashed lines whose relative distance to any of the three major associations is larger than $d_{\rm min}=1\mpc$ (see text).}
\label{fig:xyz_3}
\end{figure*}
Fig.~\ref{fig:cloud_ol_h} shows posterior samplings for the cosmological parameters $\Omega_\Lambda$ and $h$ calculated from the 50 mock data sets shown in Fig.~\ref{fig:err}. This plot includes a few results of note. First, the parameters $\Omega_\Lambda$ and $h$ exhibit a strong degeneracy which roughly runs parallel to the isochrone $t_0=t_0(\Omega_\Lambda, h)=13.46$ Gyr (dashed line) derived from Equation~(\ref{eq:age}) in a flat Universe. Yet, the degenerate solutions are still consistent with the true values (red dots) for both `central' and `pair' N-body models. Note also that the potential quadrupole does not introduce a bias on the cosmological parameters in spite of the relatively large magnitude of the transverse velocity component added to the velocity vector of all test particles ($v_t=100\kms$). Indeed, the most distant objects in the galaxy sample considered here $(r\sim 3\mpc$) put the strongest constraints on $\Omega_\Lambda$ and $h$. As shown in Fig.~(\ref{fig:mock}) the effects of (unknown) peculiar velocities tend to be negligible on these scales.

As expected from the analytical estimates obtained in \S\ref{sec:de}, we find that the constraints on the cosmological constant are considerably weaker than those on $h$. These tests suggest that the kinematics of nearby ($r\lesssim 3\mpc$) galaxies can be used to put meaningful bounds on the Hubble constant {\it only if} prior bounds on $\Omega_\Lambda$ are incorporated in the analysis. We shall return to this point in Section~\ref{sec:results}.

\section{Observational data}\label{sec:obs}
Comparison between model and observations is done through the likelihood function built in \S\ref{sec:like}, which incorporates measurements of distances and radial velocities of individual galaxies. Recently, two major catalogues of nearby galaxies have been made publicly available: McConnachie (2012), which provides a detailed description of the properties of dwarfs at heliocentric distances $D\le 3 \mpc$, and Karachentsev et al. (2013) who have compiled data for 869 galaxies within $11 \mpc$ from the sun. Both data sets provide equivalent information within the range of overlap.

Fig.~\ref{fig:xyz_3} shows the spatial location of galaxies within a $5 \mpc$ volume. For simplicity we have aligned the coordinate system so that the axis joining the Milky Way and M31 is the X-axis. 
Note that the local Universe is considerably more rich in substructures than the restricted N-body experiments built in \S\ref{sec:num}. Clearly, the idealized initial conditions of our N-body models, which are based on a homogeneous \& isotropic Universe, cannot reproduce the complex spectrum of over-densities that drive the formation of structures in the local volume. In particular, three prominent associations (or clusters) of galaxies stand out at $r\gtrsim 3\mpc$ (outer green dotted line): IC 342/Maffei-I (Karachentsev et al. 2003), M81 (Karachentsev et al. 2002; Chiboucas et al. 2013) and Centaurus A/M83 (Karachentsev et al. 2007), which we mark with blue asterisks, squares and triangles, respectively. The middle and right panels show glimpses of an even higher level of the hierarchical galaxy-formation ladder, as many of the visible structures lie on a vast plane that connects to the Virgo cluster and its filamentary network (Tully \& Fisher 1987).

\begin{figure}
 \includegraphics[width=84mm]{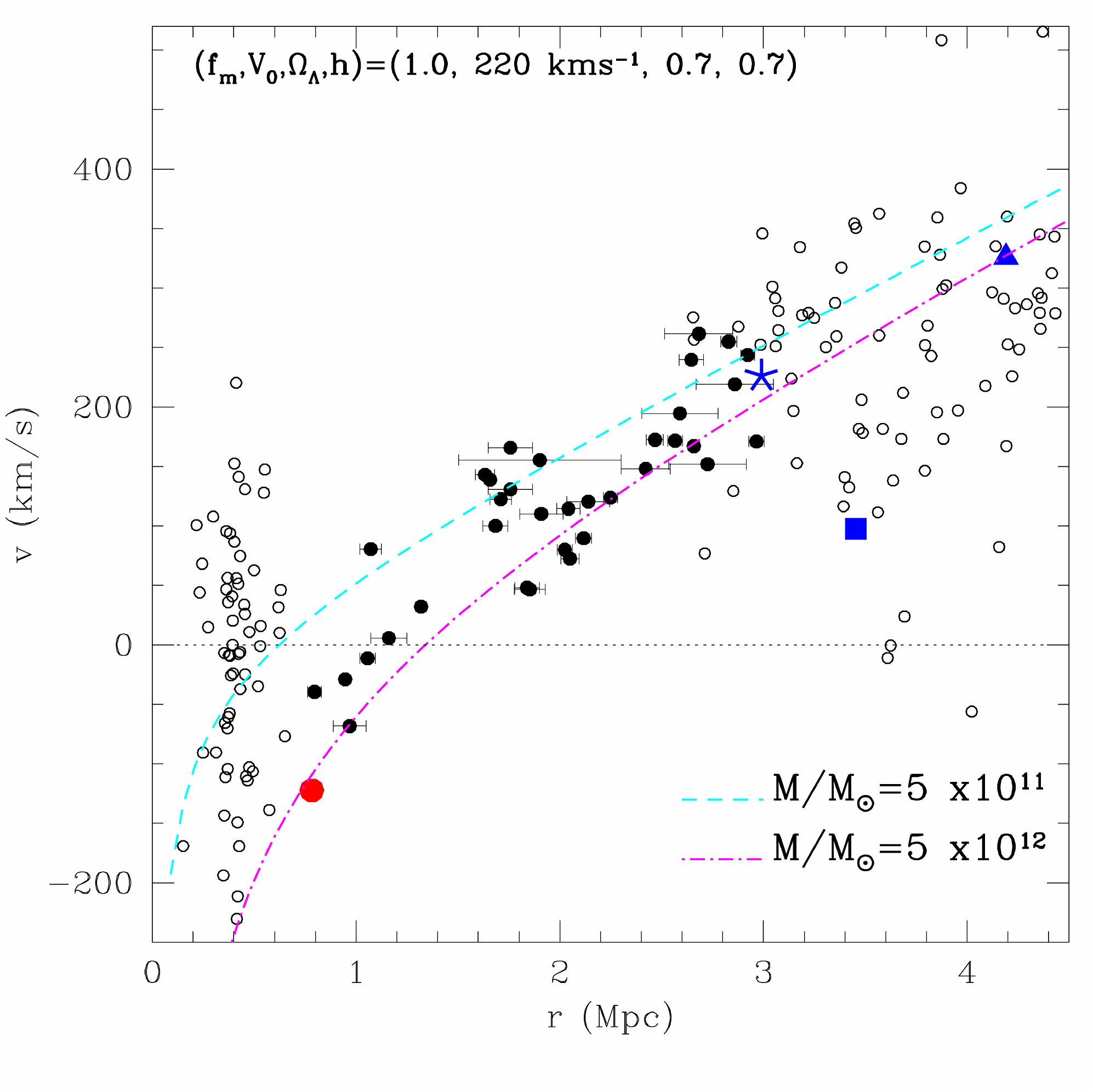}
\caption{Phase-space location of the galaxies plotted in Fig.~\ref{fig:xyz_3}. Black solid dots denote galaxies incorporated in our Bayesian analysis. A red dot marks the current separation and relative velocity between the Milky Way and M31 using $V_0=220\kms$. For reference we also plot two isochrone lines for $M=5\times 10^{11}{\rm M}_\odot$ and $M=5\times 10^{12}{\rm M}_\odot$. Note that the zero-velocity radius of the Local Group is located at $r_0\sim 1\mpc$. Regions of large velocity scatter point toward the presence of a large number of satellite galaxies.}
\label{fig:rv}
\end{figure}

We identify the Milky Way and M31 (red dots) as well as the overdensities that surround them. These correspond to gravitationally-bound `satellite' galaxies, which tend to be located at barycentric distances $r\sim d$ (inner green dotted line). Galaxies that are gravitationally bound to our Galaxy, Andromeda, or any of the external associations/clusters move on orbits that strongly deviate from the model assumptions on which the timing argument rests. To illustrate this point we plot in Fig.~\ref{fig:rv} the distance and radial velocity with respect to the Local Group barycentre of the galaxies shown in Fig.~\ref{fig:xyz_3}. Notice the large velocity scatter shown by satellites in the neighbourhood of the main galaxies (see \S\ref{sec:like}). In particular the velocity dispersion about the bulk flow increases noticeably at $r\lesssim d\approx  0.78\mpc$ and $r\gtrsim 3\mpc$. 

The impact of external perturbers such as IC 342, M81 and Centaurus A on our fits can be strongly suppressed by excluding galaxies in the vicinity of those systems\footnote{An alternative route can be pursued by adding additional terms in the equations of motion to account for the gravitational perturbations induced by the structures surrounding the Local Group (e.g. Mohayaee \& Tully 2005; Courtois et al. 2012). Such undertaking, however, introduces a number of complexities that go beyond the scope of this paper.}. Accordingly, our data set only include galaxies whose distance to any of the three major associations is larger than a given $d_{\rm min}$. In addition, we also impose a distance cut, $r_{\rm max}$, in our selection criteria, which is motivated by the decreasing accuracy of the timing argument at distances where the contribution of the Local Group to the local gravitational field becomes negligible.
 
Black solid dots in Figs.~\ref{fig:xyz_3} and~\ref{fig:rv} show galaxies that obey the above criteria for $d_{\rm min}=1\mpc$ and $r_{\rm max}=3\mpc$ (see also Table~\ref{tab:obs}). Although these hard cuts are put {\it ad hoc}, we have checked that any combination within the range $1.0\le d_{\rm min}/\mpc\le 2.0$ and $2.7\le r_{\rm max}/\mpc\le 3.0$ yields a similar fit to the model parameters, indicating that the results discussed in Section~\ref{sec:results} are not overly sensitive to the choice of $d_{\rm min}$ and $r_{\rm max}$.


In spite of the small number of dwarfs that match the above conditions ($N_{\rm sample}\sim 30$), the analysis of mock data in Section~\ref{sec:like} suggests that the set size may be sufficiently large to provide meaningful constraints on the individual masses of the Milky Way and M31, the circular velocity of the Milky Way at the solar radius, as well as on the cosmological parameters $\Omega_\Lambda$ and $h$. We explore this possibility below.

\begin{table*}
\centering
\renewcommand{\tabcolsep}{0.3cm}
\renewcommand{\arraystretch}{1.5}
\begin{tabular}{| l c c c c c c c c c c|}
\hline
\hline
Name & $l$  & $b$ & $D$ & 
$\epsilon_D$ & $V_h$ & $\epsilon_V$ & $d_{\rm IC342}$ & $d_{\rm M81}$ & $d_{\rm CenA}$&  Ref. \\
& [deg] & [deg] & [Mpc]   &[Mpc]  & [$\kms$]  & [$\kms$]  &[Mpc]  &[Mpc]  &[Mpc] & \\

\hline
Andromeda (M31)        &     121.2 &     -21.6 &    0.783 &      0.025 &    -300.0 &       4.0 & 2.71 & 3.35 & 4.57 & (1)\\
Leo A                  &     196.9 &      52.4 &    0.798 &      0.044 &      22.3 &       2.9 & 3.02 & 3.00 & 3.85 & (1)\\
Tucana                 &     322.9 &     -47.4 &    0.887 &      0.049 &     194.0 &       4.3 & 4.04 & 4.47 & 3.56 & (1)\\
WLM                    &      75.9 &     -73.6 &    0.933 &      0.034 &    -130.0 &       1.0 & 3.47 & 4.18 & 4.32 & (1)\\
Sagittarius dIrr       &      21.1 &     -16.3 &    1.067 &      0.088 &     -78.5 &       1.0 & 3.92 & 4.29 & 3.74 & (1)\\
Aquarius (DDO 210)     &      34.0 &     -31.3 &    1.072 &      0.039 &    -140.7 &       2.5 & 3.76 & 4.47 & 3.56 & (1)\\
NGC 3109               &     262.1 &      23.1 &    1.300 &      0.048 &     403.0 &       2.0 & 4.03 & 3.94 & 3.00 & (1)\\
Antlia                 &     263.1 &      22.3 &    1.349 &      0.062 &     362.0 &       2.0 & 4.09 & 3.98 & 2.97 & (1)\\
Andromeda XVIII        &     113.9 &     -16.9 &    1.355 &      0.081 &    -326.2 &       2.7 & 2.35 & 3.23 & 5.12 &(1)\\
UGC 4879               &     164.7 &      42.9 &    1.361 &      0.025 &     -29.1 &       1.3 & 2.41 & 2.33 & 4.44 & (1),(3)\\
Sextans B              &     233.2 &      43.8 &    1.426 &      0.020 &     304.0 &       1.0 & 3.51 & 3.23 & 3.50 & (1)\\
Sextans A              &     246.1 &      39.9 &    1.432 &      0.053 &     324.0 &       2.0 & 3.75 & 3.48 & 3.26 & (1) \\
HIZSS 3[A]             &     217.7 &       0.1 &    1.675 &      0.108 &     288.0 &       2.5 & 3.42 & 3.67 & 4.19 & (1)\\
HIZSS 3[B]             &     217.7 &       0.1 &    1.675 &      0.108 &     322.6 &       1.4 & 3.42 & 3.67 & 4.19 & (1)\\
Leo P                  &     219.6 &      54.4 &    1.720 &      0.400 &     264.0 &       2.0 & 3.34 & 2.85 & 3.72 & (4),(5)\\
KKR 25                 &      83.9 &      44.4 &    1.905 &      0.061 &    -139.5 &       1.0 & 2.78 & 2.53 & 4.64 & (1)\\
NGC 55                 &     332.9 &     -75.7 &    1.932 &      0.107 &     129.0 &       2.0 & 4.46 & 5.30 & 4.45 & (1)\\
IC 5152                &     343.9 &     -50.2 &    1.950 &      0.045 &     122.0 &       2.0 & 4.88 & 5.49 & 3.83 & (1)\\
ESO 294- G 010         &     320.4 &     -74.4 &    2.032 &      0.037 &     117.0 &       5.0 & 4.57 & 5.41 & 4.44 & (1)\\
NGC 300                &     299.2 &     -79.4 &    2.080 &      0.057 &     146.0 &       2.0 & 4.48 & 5.37 & 4.61 & (1)\\
GR 8                   &     310.7 &      77.0 &    2.178 &      0.120 &     213.9 &       2.5 & 4.03 & 3.21 & 3.21 & (1)\\
KKR 3 (KK 230)         &      63.7 &      72.0 &    2.188 &      0.121 &      63.3 &       1.8 & 3.44 & 2.69 & 3.98 & (1)\\
UKS 2323-326 (UGCA 438)&      11.9 &     -70.9 &    2.208 &      0.092 &      62.0 &       5.0 & 4.57 & 5.47 & 4.69 & (1)\\
IC 3104                &     301.4 &     -17.0 &    2.270 &      0.188 &     429.0 &       4.0 & 5.51 & 5.68 & 2.42 & (1)\\
UGC 9128 (DDO 187)     &      25.6 &      70.5 &    2.291 &      0.042 &     152.0 &       1.0 & 3.93 & 3.15 & 3.59 & (1)\\
IC 4662                &     328.5 &     -17.8 &    2.443 &      0.191 &     302.0 &       3.0 & 5.71 & 5.92 & 2.56 & (1)\\
KKH 98                 &     109.1 &     -22.4 &    2.523 &      0.105 &    -136.9 &       1.0 & 2.28 & 3.63 & 6.24 & (1)\\
DDO 125                &     137.8 &      72.9 &    2.582 &      0.059 &     194.9 &       0.2 & 3.11 & 1.97 & 4.50 & (1)\\
UGC 8508               &     111.1 &      61.3 &    2.582 &      0.036 &      56.0 &       5.0 & 2.77 & 1.78 & 4.88 & (1)\\
KKH 86                 &     339.0 &      62.6 &    2.582 &      0.190 &     287.2 &       0.7 & 4.69 & 3.87 & 2.81 & (1)\\
DDO 99                 &     166.2 &      72.7 &    2.594 &      0.167 &     251.0 &       4.0 & 3.20 & 2.05 & 4.40 & (1)\\
DDO 190                &      82.0 &      64.5 &    2.793 &      0.039 &     150.0 &       4.0 & 3.36 & 2.37 & 4.66 & (1)\\
NGC 4163               &     163.2 &      77.7 &    2.858 &      0.039 &     165.0 &       5.0 & 3.48 & 2.21 & 4.38 & (1)\\
NGC 404                &     127.1 &     -27.0 &    3.060 &      0.370 &     -48.0 &       9.0 & 2.14 & 3.81 & 6.83 & (2)\\
 \hline
\end{tabular}
\caption[]{Heliocentric coordinates \& radial velocities of dwarf galaxies within a radial range $0.8\le r/\mpc \le 3$ from the Local Group barycentre and with separations to IC 342, M81 and Centaurus A larger than $1\mpc$ (see text). Data taken from (1) McConnachie et al. (2012); (2) Karachentsev et al. (2002); (3) Kirby et al. (2012); (4) Giovanelli et al. (2013); (5) McQuinn et al. (2013). Relative distances to the major associations in the vicinity of the Local Group are calculated using a fiducial $f_{\rm m}=1$ and the following heliocentric positions, $(D,l,b)_{\rm IC348}=(3.3\mpc,138.17^\circ,+10.58^\circ)$, $(D,l,b)_{\rm M81}=(3.6\mpc,142.09^\circ,+40.90^\circ)$ and $(D,l,b)_{\rm CenA}=(3.8\mpc,309.52^\circ,+19.42^\circ)$. }
\label{tab:obs}
\end{table*}

\begin{table}
\centering
\renewcommand{\tabcolsep}{0.3cm}
\renewcommand{\arraystretch}{1.5}
\begin{tabular}{| l l l |}
\hline
\hline
Model parameters & flat priors & {\it Planck} prior on $\Omega_\Lambda$  \\
\hline
$M/(10^{12}{\rm M}_\odot)$      & $2.3_{-0.7(-1.2)}^{+0.7(+1.7)}$      & $2.3_{-0.7(-1.2)}^{+0.7(+1.7)}$  \\
$f_{\rm m}$                     & $0.54_{-0.17(-0.30)}^{+0.23(+0.60)}$ & $0.54_{-0.16(-0.30)}^{+0.24(+0.60)}$ \\
$V_0/\kms$                     & $245_{-23(-45)}^{+23(+47)}$          & $245_{-23(-45)}^{+23(+51)}$  \\
$\sigma_m/\kms$                & $35_{-4(-8)}^{+5(+11)}$             & $35_{-4(-8)}^{+6(+12)}$  \\
$\Omega_\Lambda$                & $0.54_{-0.35(-0.51)}^{+0.32(+0.44)}$ & $0.69_{-0.02(-0.04)}^{+0.02(+0.04)}$ \\
$h$                            & $0.64_{-0.07(-0.12)}^{+0.10(+0.20)}$ & $0.67_{-0.04(-0.09)}^{+0.04(+0.09)}$ \\
\hline
Derived quantities &  & \\
\hline
$|v_\odot|/\kms$                  & $312_{-11(-22)}^{+11(+23)}$         &  $313_{-11(-22)}^{+11(+23)}$\\
$l_\odot$ (deg.)                  & $93.9_{-1.6(-3.0)}^{+1.8(+3.6)}$    &  $93.9_{-1.5(-2.9)}^{+1.7(+3.5)}$\\
$b_\odot$ (deg.)                  & $-3.2_{-1.3(-2.7)}^{+1.2(+2.3)}$    & $-3.2_{-1.3(-2.7)}^{+1.2(+2.2)}$\\
$M_{\rm G}/(10^{12}{\rm M}_\odot)$ & $0.8_{-0.3(-0.5)}^{+0.4(+0.9)}$      & $0.8_{-0.3(-0.5)}^{+0.4(+0.9)}$\\
$M_{\rm A}/(10^{12}{\rm M}_\odot)$ & $1.5_{-0.4(-0.8)}^{+0.5(+1.2)}$      & $1.5_{-0.4(-0.8)}^{+0.5(+1.1)}$ \\ 
$t_0/{\rm Gyr}$                  & $13.2_{-1.4(-2.4)}^{+2.9(+8.5)}$     &  $13.8_{-0.8(-1.7)}^{+1.0(+2.1)}$  \\ 
$\sigma_{\rm H}/\kms$             & $50_{-4(-8)}^{+4(+8)}$              &  $50_{-4(-8)}^{+4(+8)}$  \\ 
\hline
\end{tabular}
\caption[]{Constraints on model parameters using flat priors (Table~\ref{tab:priors}), and a Gaussian prior on the fractional vacuum energy density based on {\it Planck} data ($\Omega_\Lambda=0.686\pm 0.020$). Error bars enclose the central 68\% (95\%) of area under the marginalized 1D posterior probability distribution functions shown in Fig.~\ref{fig:grid}. }
\label{tab:results}
\end{table}

\section{Results}\label{sec:results}
\begin{figure*}
\vspace*{0cm}\includegraphics[width=162mm]{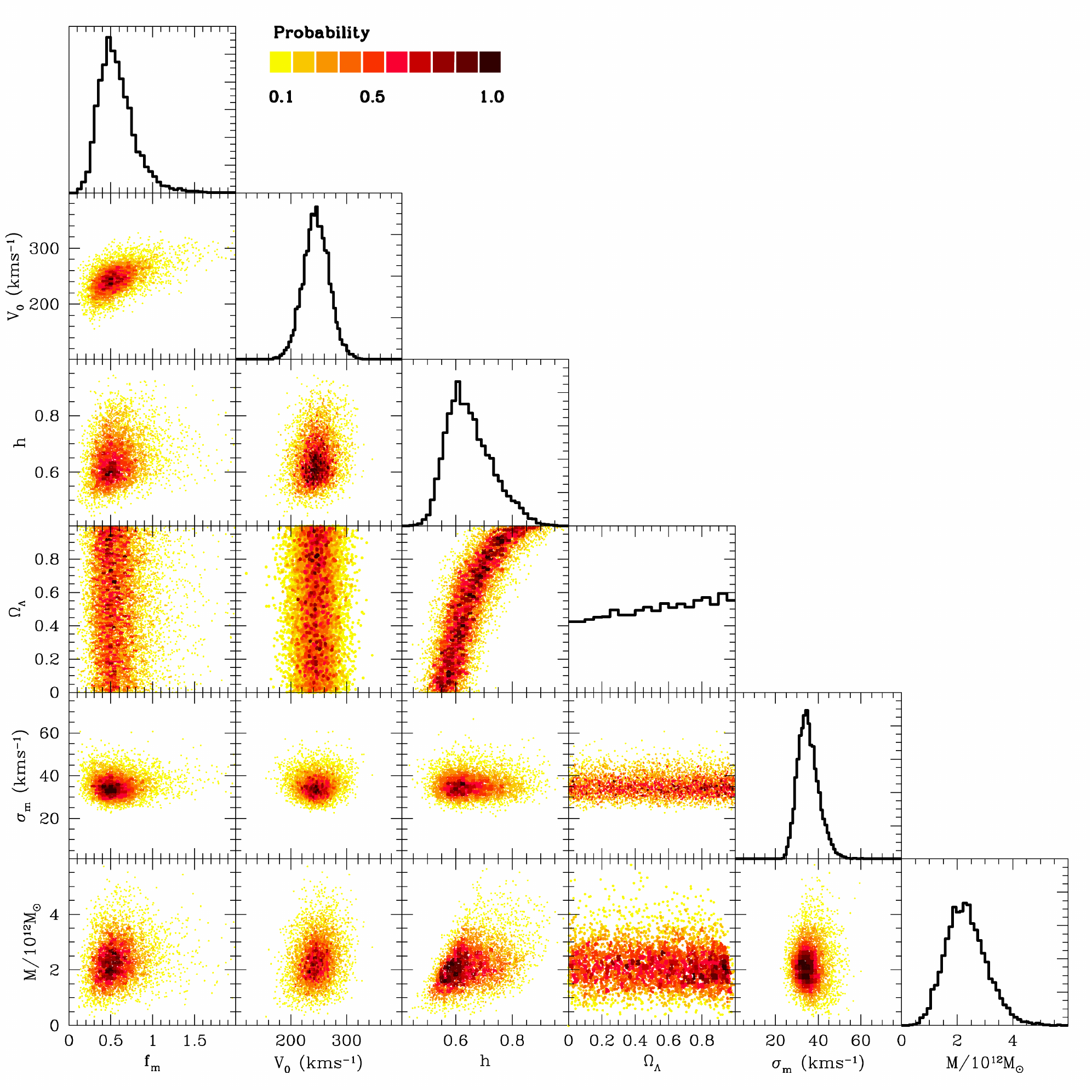}
\caption{Posterior distributions for our model parameters. Fits include galaxies within the radial range $0.8\le r/\mpc\le 3.0$ and a separation to IC 342, M81 and Centaurus A larger than $d_{\rm min}=1\mpc$ (see Table~\ref{tab:obs}). }
\label{fig:grid}
\end{figure*}

We now apply our method to galaxies located within the radial range $0.8\mpc \le r\le r_{\rm max}$ and a separation to IC 342, M81 and Centaurus A larger than $d_{\rm min}=1\mpc$. The largest number of galaxies in the sample is $N_{\rm sample}=30$ for $r_{\rm max}= 3\mpc$ (see Table~\ref{tab:obs}). Fig.~\ref{fig:grid} displays posterior distributions for each model parameter as returned by {\sc MultiNest}. Table~\ref{tab:results} lists for each parameter the median value from the posterior PDF, with error bars indicating the interval that encloses the central 68\% and 95\% of values. 

\subsection{Solar apex}
A correct estimation of the motion of the Sun relative to the other members of the Local Group is a key aspect of our study. The conversion between helio- and Local Group-centric coordinates (see Appendix A) requires an understanding of the relative motion between the sun and the Milky Way, as well as between the Milky Way and the Local Group barycentre. Both steps introduce a non-negligible degree of uncertainty in our models. 

Although the motion of the sun has been studied for many decades, this is the first attempt to measure the location of its apex by modelling the kinematics of individual Local Group members while simultaneously dealing with uncertainties in the circular velocity of the sun as well as on the relative mass between the Milky Way and M31. 

It is thus reassuring that the coordinates listed in Table~\ref{tab:results} agree very well with previous measurements. For example, Karachentsev \& Makarov (1996) minimize scatter in the distribution of radial velocities with respect to the bulk flow, $v=H r$, where $H$ is a free parameter related to the (local) expansion of the Universe (see also Appendix B). These authors find $(v,l,b)_\odot=(316\pm 5\kms, 93^\circ\pm 2^\circ, -4^\circ\pm 2^\circ)$, which is consistent with our measurement within one-sigma uncertainties. Similarly, Courteau \& van den Bergh (1999) assume that the velocity distribution of nearby galaxies with respect to the Hubble flow is Maxwellian. These authors argue that if the velocity distribution is invariant under spatial translations, the true solar motion is the one that minimizes the velocity dispersion of their models\footnote{Interested readers can find a formal proof of this argument in Pe\~narrubia, Koposov \& Walker (2012).}. Their result $(v,l,b)_\odot=(306\pm 18\kms, 99^\circ\pm 5^\circ, -3^\circ\pm 4^\circ)$ also agrees with our measurement at one-sigma confidence level. 

\subsection{The Sun's transverse motion}\label{sec:v0}
The circular velocity of the Milky Way at the solar radius, $V_0\equiv V_c(R_\odot)$, is a crucial parameter for the derivation of the solar apex. Unfortunately, the wide range of values reported in the literature (e.g. Bhattacharjee et al. 2013) cannot be explained by the quoted errors of individual measurements, which may be indicative of systematic biases in some of the published methods. Given that a wrong choice of $V_0$ propagates through our whole analysis, here we opt for {\it not} imposing an external constraint on its value. In a Bayesian framework this is equivalent to adopting a 'diffuse' or 'uninformative' prior (see Table~\ref{tab:priors}). Note that by fitting $V_0$ simultaneously with the rest of parameters we are effectively incorporating the uncertainty in the value of $V_0$ into the joint posterior distributions of {\it all} measured quantities.  

Fig.~\ref{fig:grid} shows that the kinematics of nearby galaxies can be used to put meaningful constraints on the value of $V_0$ if we adopt prior information on the motion of the Local Standard of Rest (LSR; see Appendix A). Indeed, the small covariance of $V_0$ with the rest of parameters is at the core of the relative narrowness of the confidence intervals given in Table~\ref{tab:results}. 

Adopting the LSR velocity vector measured by Sch\"onrich et al. (2010), the median value returned from our Bayesian fits is $V_0=245\pm 23\kms$ at a 68\% confidence level. Although these bounds lie above the value of $220\kms$ adopted by the IAU (Kerr \& Lynden-Bell 1986), our measurement appears in excellent agreement with the recent estimates of McMillan (2011), who finds $V_0=239\pm 5\kms$ via modelling the kinematics of stars in the Milky Way disc, and with Bovy et al. (2009) who obtain $V_0=246\pm 30\kms$ from trigonometric parallaxes. 
It is interesting to note that Arp (1986) found $V_0=239\pm 17\kms$ by minimizing the velocity dispersion of nearby galaxies about the bulk flow, an argument very similar to the one exposed in Appendix B.
Recently, Sch\"onrich (2012) has devised a model-independent method for measuring $V_0$ using the position and velocities of kinematically hot stars in the solar neighbourhood, which yields $V_0=238\pm 9\kms$, which also falls within the confidence interval of our fits. 
In contrast, Bovy et al. (2011) find $V_0=218\pm 6\kms$ using data from the APOGEE spectroscopic survey. However, these authors also detect an offset between the Sun's rotational velocity with respect to the Local Standard of Rest of $\approx 22\kms$, which is a factor $\sim 2$ larger than the one measured by Sch\"onrich et al. (2010). Accounting for this offset seems to reconcile their value with the one listed in Table~\ref{tab:results}.
Finally, if we add the LSR azimuthal component to the value of $V_0$ we find that the transverse velocity component of the sun with respect to the Milky Way is $V_\phi\approx 257\kms$, which is in good agreement with the value derived by Reid et al. (2009), $V_\phi= 254\pm 16 \kms$, from proper motions of masers in the Milky Way.


\subsection{The (very) local value of $H_0$}
Fig.~\ref{fig:grid} shows a strong covariance between the cosmological parameters of our analysis, $\Omega_\Lambda$ and $h$. The tests run in \S\ref{sec:num} suggest that the correlation follows approximately the isochrone $t_0=t_0(\Omega_\Lambda,h)$, where $t_0$ is the age of the Universe. The strong degeneracy arises from the scant sensitivity of the kinematics of nearby galaxies to the vacuum energy term in the equations of motion (see \S\ref{sec:de}).

\begin{figure}
 \includegraphics[width=84mm]{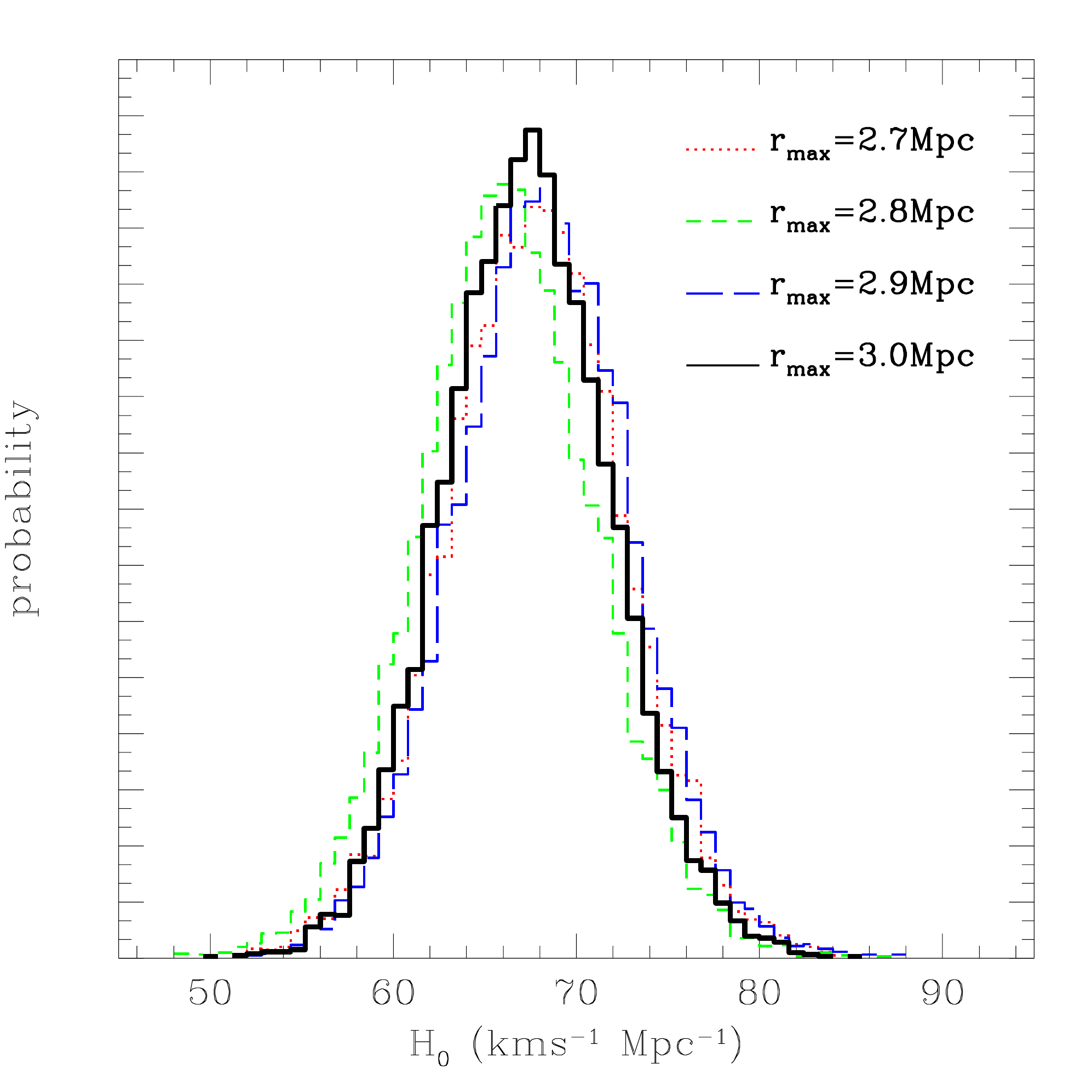}
\caption{Posterior distribution functions for the Hubble constant $H_0$ derived from four different samples of nearby galaxies with distances to the Local Group barycentre $r<r_{\rm max}$. In order to break the degeneracy between $H_0$ and $\Omega_\Lambda$ shown in Fig.~\ref{fig:grid} we adopt {\it Planck}'s measurement of the fractional vacuum density, $\Omega_\Lambda=0.686\pm 0.020$, as a Gaussian prior in our fits. Note that our constraints on $H_0$ are insensitive to the value of $r_{\rm max}\lesssim 3\mpc$. Beyond $\sim 3\mpc$ the motion of galaxies appear to be strongly perturbed by galaxy clusters in the Local Group vicinity (see Figs.~\ref{fig:xyz_3} and~\ref{fig:rv}). }
\label{fig:h}
\end{figure}

However, the shape of the covariance is such that even a modest prior on $\Omega_\Lambda$ may be sufficient to put a tight bound on the value of the Hubble constant. Indeed, Fig.~\ref{fig:h} shows that the posterior distribution function of $H_0$ returned by our analysis becomes relatively narrow once we incorporate the bounds on $\Omega_\Lambda$ obtained from the spectrum of fluctuations in the Cosmic Microwave Background (CMB) as priors in our analysis. For simplicity, we adopt a Gaussian prior on the value of the fractional vacuum density that follows the posterior distribution function derived from the analysis of the {\it Planck} data, i.e. $\Omega_\Lambda=0.686\pm 0.020$. Imposing distance cuts to the galaxy sample in the range $2.7\le r_{\rm max}/\mpc\le 3.0$ does not significantly change our constraints. 

Combination of CMB data {\it and} the dynamics of Local Group galaxies yields a local Hubble constant $H_0=67\pm 5\kms\mpc^{-1}$ at a 68\% confidence level. This result agrees very well with {\it Planck}'s constraint, $H_0=67.4\pm 1.4 \kms\mpc^{-1}$, and is also consistent at a one-sigma level with the value derived from Cepheid data, $H_0=72.5\pm 2.5\kms\mpc^{-1}$ (Efstathiou 2013 and references therein), as well as with fits to the Hubble diagram over cosmological scales, e.g. $H_0=74.4\pm 3.0\kms\mpc^{-1}$ (Tully et al. 2013); $H_0=74\pm 4\kms\mpc^{-1}$ (Sorce et al. 2013). 

The velocity dispersion of the flow in the radial range sampled by our data ($\sigma_{\rm H}=50\pm 4\kms$) is consistent with that expected galaxies embedded in large-scale walls (Aragon-Calvo et al. 2011).

 
Combination of {\it Planck}'s constraint on the vacuum energy density with our measurement of the Hubble constant gives an estimate of the Universe's age, $t_0=13.8_{-0.9}^{+1.0}$Gyr (see Table~\ref{tab:results}).

Alternatively, one can use the age of the Universe obtained by {\it Planck}, $t_0=13.813\pm 0.058$Gyr, as a prior in our models. In this case the parameter $\Omega_\Lambda h^2$ is entirely constrained from the effects of the pressure term on the dynamics of nearby galaxies. Although note shown here, we obtain bounds on $\Omega_\Lambda h^2$ that are in agreement with those given in Table~\ref{tab:results}, highlighting the consistency between the Hubble constant derived from local dynamics and its cosmological value.


\subsection{Milky Way and Andromeda masses}\label{sec:mass}
The masses of the Milky Way and Andromeda follow directly from the bounds on $M$ and $f_{\rm m}$. The measurement rests on the assumption that both systems account for the entire mass of the Local Group, i.e. $M=M_{\rm G}+M_{\rm A}$, which appears reasonable given that the third brightest galaxy in our vicinity, M33, rotates with a relatively low speed\footnote{Deep photometric surveys of the M31-M33 system show that M33 acted on by tides (McConnachie et al. 2009), which may alter the outer rotation curve of this galaxy.}, $V_{\rm rot,M33}\approx 120 \kms$ (Corbelli \& Salucci 2000), whereas both the Milky Way and M31 have rotation velocities peaking at $\sim 250\kms$. Indeed, combination of the Tully-Fisher relation (1977) with the rotational velocity for our own Galaxy indicates that M33 accounts for a tiny fraction of the Local Group mass, $M_{\rm M33}/M\sim (V_{\rm rot, M33}/V_0)^4f_{\rm m}/(1+f_{\rm m})\simeq 0.028$.

The median values of the Milky Way and Andromeda masses are $M_{\rm G}=0.8_{-0.3}^{+0.4}\times 10^{12}{\rm M}_\odot$ and $M_{\rm A}=1.5_{-0.4}^{+0.5}\times 10^{12}{\rm M}_\odot$ at a 68\% level (see Table~\ref{tab:results}). Models where the Milky Way is more massive than Andromeda (i.e. $f_{\rm m}\ge 1$) are ruled out with high significance ($\sim 95\%$). 

It is worth stressing that, by virtue of the large distance range covered by the sample of Local Group galaxies, the quoted values correspond to the {\it total mass} of both galaxies. In contrast, most dynamical models in the literature use kinematic tracers (e.g. gas, stars, planetary nebulae and/or globular clusters) that only populate the inner regions of galactic haloes. From Newton's theorem any mass distribution outside the limiting radius of the data has no observational effect in a spherical or elliptical system. Hence, these models must assume a density profile in order to extrapolate the inner mass bounds to radii devoid of visible tracers (e.g. the virial radius, $r_{\rm vir}\sim 260\kpc$). Unfortunately, the outer mass profile of the Milky Way remains unknown, increasing the uncertainty of the extrapolation. 
For example, Smith et al. (2007) use high-velocity stars from the RAVE survey (Steinmetz et al. 2006; Zwitter et al. 2008) to measure the local escape speed of our Galaxy. The range of values found by these authors ($498<v_{\rm esc}/\kms<608$) leads to a virial mass $M_{\rm G, vir}=0.85_{-0.29}^{+0.55}\times 10^{12}{\rm M}_\odot$ under the assumption that the Milky Way halo follows a NFW profile (Navarro et al. 1997). However, if the dark matter halo contracts as a result of dissipative processes (Mo et al. 1998) the mass estimate increases to $1.42_{-0.54}^{+1.14} \times 10^{12}{\rm M}_\odot$. 
Xue et al. (2008) extend the mass constraints to $D\lesssim 60\kpc$ by modelling the kinematics of blue horizontal branch stars (BHBs). 
Depending on whether or not the halo models are contracted the mass estimate varies between $M_{\rm G, vir}=1.2_{-0.3}^{+0.4}\times 10^{12}{\rm M}_\odot$ and $0.8_{-0.2}^{+0.2}\times 10^{12}{\rm M}_\odot$, respectively. 
This range is broadly consistent with Sakamoto et al. (2003) and Battaglia et al. (2005) who find $0.8_{-0.2}^{+1.2}\times 10^{12}{\rm M}_\odot$ and $2.5_{-1.0}^{+0.5}\times 10^{12}{\rm M}_\odot$, respectively, from a mix sample of globular cluster, giant stars and satellite galaxies; as well as with the recent estimates of Deason et al. (2012) based on Jeans modelling a sample of BHB stars out to $D\sim 90\kpc$.

In principle, satellite galaxies sample a much larger volume of the galactic halo and may provide a tighter constraint on the virial mass of the host galaxy. In practice, their poorly known orbital distribution introduces a considerable uncertainty in this type of analysis. Using the locations and kinematics of 26 Milky Way satellites Watkins et al. (2010) constrain the Milky Way mass to lie within $0.7$-$3.4\times 10^{12}{\rm M}_\odot$ depending on the assumed velocity anisotropy. Incorporating the proper motions of 6 satellites into the analysis narrows the mass range to $1.4\pm 0.3\times 10^{12}{\rm M}_\odot$. Recently, Barber et al. (2013) have carried a comparison between the orbits of Milky Way satellites with known proper motions and the eccentricity distribution observed in the Aquarius N-body simulations\footnote{The Aquarius project consists of six dark matter-only realizations of a $\sim 10^{12}{\rm M}_\odot$ halo which do not account for the enhanced disruption rate of satellites moving on highly eccentric orbits under the presence of a disc component (D'Onghia et al. 2010, Pe\~narrubia et al. 2010)}. From this exercise they find that the Milky Way mass lies $0.6$-$3.1\times 10^{12}{\rm M}_\odot$ with a best-fit value of $\sim 1.1\times 10^{12}{\rm M}_\odot$. 

Our measurement of M31 mass, $M_{\rm A}=(1.5\pm 0.3)\times 10^{12}{\rm M}_\odot$, is in agreement with existing estimates. Seigar et al. (2008), who improve on Kyplin et al. (2002) analysis of the HI rotation curve using Spitzer 3.6$-\mu$m data, adopt an adiabatically contracted NFW halo profile, finding $M_{\rm A, vir}=(0.82\pm 0.02)\times 10^{12}{\rm M}_\odot$, which is broadly consistent with the value of $\sim 0.77 \times 10^{12} {\rm M}_\odot$ inferred by Geehan et al. (2006) and with the estimate of $(0.37 - 2.1)\times 10^{12}{\rm M}_\odot$ derived from satellite kinematics (C\^{o}t\'e et al. 2000). It is also very close to the lower limit of $0.9 \times 10^{12} {\rm M}_\odot$ obtained from kinematics of halo stars (Chapman et al. 2006). From the kinematics of 23 satellites Watkins et al. (2010) find that a mass in the range $0.85$-$1.6\times 10^{12}{\rm M}_\odot$, concluding that the large uncertainty arising from the unknown orbital anisotropy prevents them to determine which of the two galaxies is actually the more massive. 
Using line-of-sight velocities for a sample of globular clusters in the stellar halo of M31 Veljanoski et al. (2013) estimate a mass $1.2$-$1.5 \times 10^{12} {\rm M}_\odot$, although the apparent association of many of these clusters to stellar streams (Mackey et al. 2010) may have some impact on the result. 
Stellar streams provide an independent constraint on the potential of M31. Fitting the kinematics the Giant Stream Ibata et al. (2004) and Fardal et al. (2013) 
derive a virial mass of $(1.0\pm 0.5)\times 10^{12}{\rm M}_\odot$ and $(2.0\pm 0.5)\times 10^{12}{\rm M}_\odot$, respectively. 

Overall we find that the combined dynamical masses of the Milky Way and Andromeda published in the literature roughly match the Local Group mass derived from our analysis of the local Hubble flow. We discuss the implications of this result below.

\begin{figure}
 \includegraphics[width=86mm]{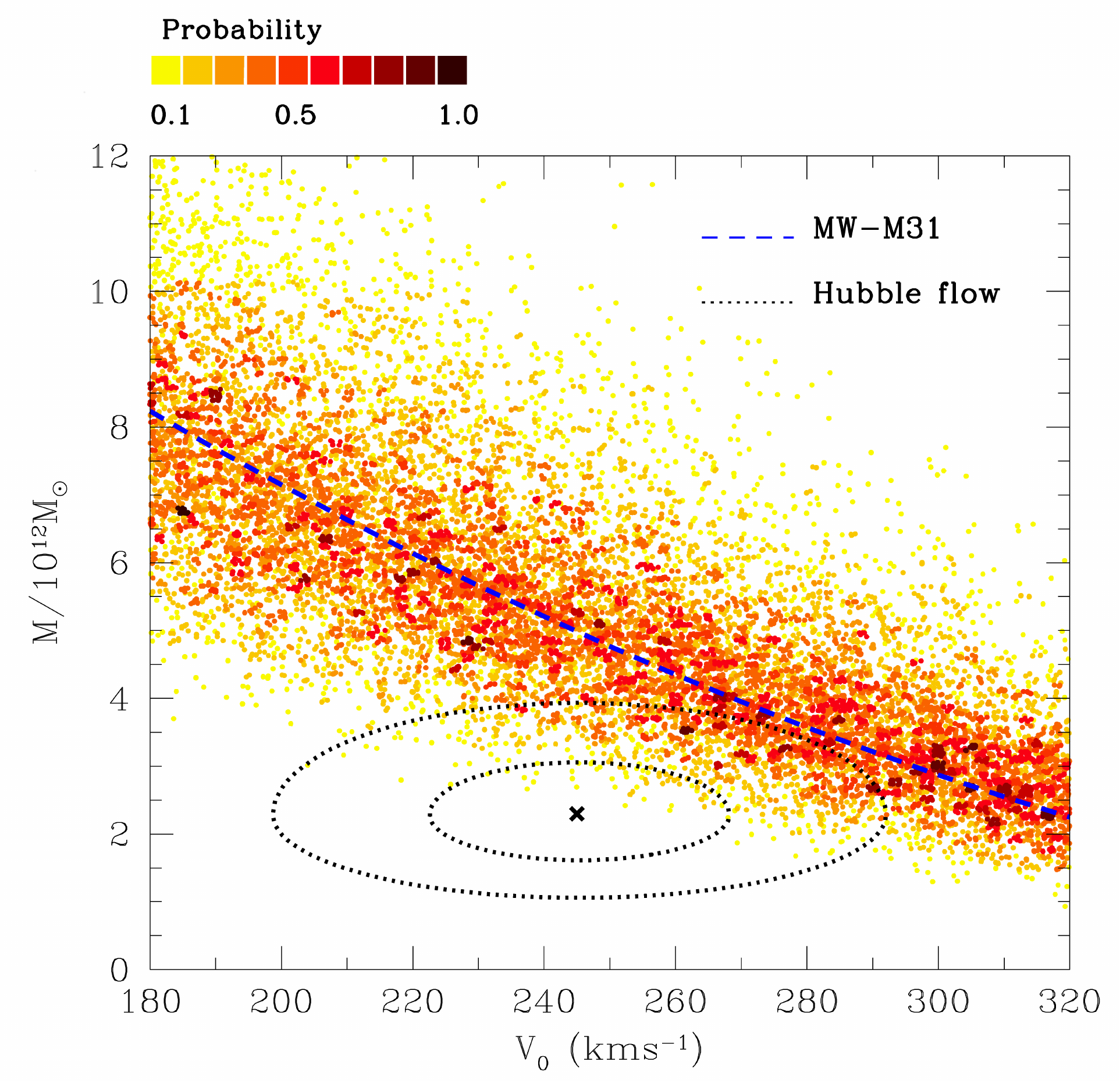}
\caption{Constraints on the Local Group mass as a function of the circular velocity of the Milky Way at the solar radius ($V_0$) using the relative motion between the Milky Way and M31 (the so-called `timing argument') and the kinematics of nearby galaxies (`Hubble flow'). Colour-coded dots sample the posterior distribution derived from the relative motion between the Milky Way and M31 using $h=0.67\pm 0.05$ and {\it Planck}'s prior on the fractional vacuum density, $\Omega_\Lambda=0.686\pm 0.020$. For reference, a blue dashed line marks the solution to Equation~(\ref{eq:timing}) with $t_0=13.8{\rm Gyr}$, $r=d(t_0)=0.783\mpc$ and $V_{h,{\rm A}}=-300\kms$. 
Dotted lines mark 68\% and 95\% confidence intervals derived from our fits to the local Hubble flow (see Fig.~\ref{fig:grid}) using {\it uninformative} priors on the cosmological parameters. The discrepancy between both methods eases by setting $V_0$ above the IAU concordance value. }
\label{fig:v0m}
\end{figure}

\section{Discussion: Missing mass in the Local Group?}\label{sec:discussion}
\subsection{Timing argument versus Hubble flow}
The Local Group mass derived from the timing argument ($\sim 5\times 10^{12}{\rm M}_\odot$, see Fig.~\ref{fig:rv}; also Li \& White 2008; van der Marel et al. 2012a,b; Yepes et al. 2013; Partridge et al. 2013) is considerably larger than the combined masses of the Milky Way and M31 ($\sim 2\times 10^{12}{\rm M}_\odot$). The existence of large amounts of `missing' mass in the Local Group with no visible counterpart poses a difficult problem to current galaxy formation models. 

Interestingly, the mass derived from the local flow, $M= 2.3\pm 0.7 \times 10^{12}{\rm M}_\odot$, is also a factor $\sim 2-3$ lower than the value suggested by the timing argument (see Section~\ref{sec:mass}). This mismacth becomes the more intriguing if we take into account that both estimates rest on Equation~(\ref{eq:kep}). It is worth discussing a number of mechanisms that could potentially bias the above measurements.

Let us start with the mass estimate derived from the local Hubble flow. In Section~(\ref{sec:quadeff}) we show that neglecting the quadrupole term in the equations of motion tends to overestimate the Local Group mass, a bias that grows under the presence of large peculiar motions in the synthetic data sets. Hence, correcting for this effect would widen the discrepancy with the timing argument even further. Also, the low velocity dispersion of the local Hubble flow, $\sigma_{\rm H}=50\pm 4\kms$ (see Table~\ref{tab:results}), suggests that the overall magnitude of the peculiar motions is small. In this case our tests indicate that neglecting the quadrupole term in Equation~(\ref{eq:kep}) leads to a minor bias in the mass estimate (see Section~\ref{sec:quadeff}; ``pair'' models).

The hierarchical mass growth of the Local Group also has little impact on the observed Hubble flow. In Section~(\ref{sec:massev}) we show that ignoring the time-dependence of the Local Group potential in Equation~(\ref{eq:kep}) tends to underestimate the Local Group mass, but the bias is so small that only a very rapid growth of the Milky Way and Andromeda masses would have a measurable effect on the observed kinematics of nearby galaxies.  

It thus appears more simple to envision mechanisms that could potentially affect the mass obtained from the timing argument. In particular, it is worth following up the results of Van der Marel et al. (2012b), who observe a strong dependence between the mass derived from the relative motion between the Milky Way and M31 and the circular velocity of the Milky Way at the solar radius ($V_0$). Combination of Equations~(\ref{eq:vd}) and~(\ref{eq:v31}) yields 
\begin{eqnarray}\label{eq:timing}
M\approx \frac{0.83d}{G}\bigg[(1.2+0.31\Omega_\Lambda)\frac{d}{t_0}-V_{h,{\rm A}} - {\bf v}_{\rm LSR}\cdot \hat {\bf r}_{\rm A} \\ \nonumber
-V_0\sin(l_{\rm A})\cos(b_{\rm A})\bigg]^2;
\end{eqnarray}
where ${\bf v}_{\rm LSR}$ is the velocity vector of the Local Standard of Rest (see Appendix A). 

Equation~(\ref{eq:timing}) shows two points of interest. First, ignoring the cosmological constant term in the equations of motion {\it lowers} the Local Group mass by a small factor $\sim 0.55\Omega_\Lambda d^{3/2}/(\sqrt{GM}t_0)\sim 0.1$, which is in good agreement with the findings of Partridge et al. (2013). 
Second, the mass suggested by the timing argument is strongly sensitive to the value of $V_0$. In particular, the minus sign in front of 
this parameter and the fact that $\sin(l_{\rm A})\cos(b_{\rm A})>0$ imply that the estimated Local Group mass drops if the circular velocity at the solar radius lies above the standard IAU value. 

Fig.~\ref{fig:v0m} illustrates this point in more detail. The blue dashed line shows the relation implied by Equation~(\ref{eq:timing}). Colour-coded dots sample the posterior distributions on $M$ and $V_0$, which reflect the uncertainty in the value of $d$ and $V_{h,{\rm A}}$, as well as in the cosmological parameters. For a better comparison with the results of previous Sections we adopt Gaussian priors on these parameters, with $H_0=67\pm 5\kms\mpc^{-1}$ and $\Omega_\Lambda=0.686\pm 0.020$. 
Comparison with the constraints derived in previous Sections indicates that the local Hubble flow provides a much tighter bound on the mass of the Local Group ($M=2.3\pm 0.7\times 10^{12}{\rm M}_\odot$; marked with dotted lines) than the timing argument, and that the amount of `missing' mass is not statistically significant once we take into account the uncertain value of $V_0$ (see \S\ref{sec:v0}). 

\subsection{The Local Group mass in a cosmological context}
The dynamical models outlined in Section~\ref{sec:eqmot} assume that the Milky Way and M31 can be treated as point masses, thus neglecting the internal distribution of matter in these galaxies. In a $\Lambda$CDM Universe galactic haloes are expected to follow a close-to-universal density profile that falls with radius as $\rho\sim r^{-3}$ (e.g. Navarro, Frenk \& White 1997). It is trivial to show that the mass profile $M(r)=4\pi\int_0^r \rho(r') r'^2\d r'$ diverges at large radii, which complicates the interpretation of the mass estimates derived from the timing argument. 

Li \& White (2008) examined this problem and found that if the ``true'' Local Group mass is taken to be the sum of the ``virial'' masses ($M_{200}$) of the two dominant galaxies, where $M_{200}$ is defined as the mass within a sphere of mean density 200 times the critical value, the ratio of the true virial mass to that estimated from the `classical' ($\Omega_\Lambda=0$ in Equation~\ref{eq:kep}) timing argument is $\sim 1.5$ (see also Yepes et al. 2013). 
Adding a cosmological constant term in the equations of motion reduces the mismatch to a factor $\sim 1.15$ (Partridge et al 2013), with a considerable scatter. 

Recent HST measurements of the transverse velocity vector between the Milky Way and M31 (van der Marel 2012a) suggest that the galaxies are approaching each other on a head-on trajectory. Gonzalez et al. (2013) have surveyed the Bolshoi simulations (Klypin et al. 2011) in an attempt to find Local Group analogues on similar orbital configurations. When comparing the timing argument mass against the true virial masses these authors find that for systems moving on nearly radial orbits the timing argument over-estimates the combined mass by a factor of $\sim 1.3$-$1.6$. Fig.~\ref{fig:v0m} shows than correcting for this bias and adopting a circular velocity $V_0\sim 245\kms$ would bring the masses derived from the timing argument and the local Hubble flow into good agreement. The result $M\sim 2\times 10^{12}M_\odot$ would be compatible with the combined dynamical masses of the Milky Way and Andromeda, thus removing the need for 'missing' mass in the Local Group.


\section{Summary}\label{sec:summary}
We have inspected the equations of motion that govern the dynamics of nearby galaxies within the $\Lambda$CDM paradigm. Using analytical arguments we show that the time-dependence of the Local Group potential has a small impact on the observed local Hubble flow.
In contrast, our analysis indicates that the orbits of galaxies in the local volume can be strongly perturbed by 
the gravitational quadrupole that arises from by the dominant masses of the Milky Way and Andromeda. In particular, the quadrupole induces an attractive force along the axis joining the main galaxies which becomes repulsive in any transverse direction. Restricted N-body experiments of the local cosmic expansion show that the infall of galaxies on to the Local Group occurs preferentially along the axis joining the Milky Way and Andromeda, leading a highly anisotropic distribution of galaxies within several Mpc of the Local Group barycentre.

We devise a Bayesian method for analyzing observations of the local Hubble flow using the timing-argument equations with a vacuum energy term. Our model fits simultaneously the Local Group mass, $M=M_{\rm G}+M_{\rm A}$, the mass ratio between our Galaxy and Andromeda, $f_{\rm m}=M_{\rm G}/M_{\rm A}$, the circular velocity of the Milky Way at the solar radius, $V_0=V_c(R_\odot)$, the reduced Hubble constant, $h$, and the fractional vacuum density, $\Omega_\Lambda$. Tests with synthetic data drawn from the restricted N-body models indicate that neglecting the potential quadrupole leads to Local Group mass estimates that can be overestimated up to $\sim 30\%$, but does not affect the constraints on the rest of model parameters.

Applying our method to published locations and radial velocities of nearby galaxies returns a mass $M=2.3\pm 0.7\times 10^{12}{\rm M}_\odot$, which is consistent with the combined dynamical masses of the Milky Way and M31 and does not require the presence of extra `missing' mass between both galaxies, and a mass ratio $f_{\rm m}=0.54^{+0.23}_{-0.17}$ which rules out dynamical models where the Milky Way is more massive than Andromeda with $\sim 95\%$ confidence. Both estimates are in very good agreement with those recently found by Diaz et al. (2014) who implement their own Bayesian modelling of the kinematics of Local Group galaxies ($r\lesssim 1.5\mpc$). The Milky Way's circular velocity at the solar radius, $V_0=245\pm 23\kms$, is slightly lower than the peak velocity of M31 (Klypin et al. 2002), lending additional support to dynamical models where $f_{\rm m}<1$.  

 The cosmological parameters $\Omega_\Lambda$ and $h$ are strongly covariant. The shape of the covariance is such that introducing the CMB prior on $\Omega_\Lambda$ is sufficient to put a tight bound on the value of the {\it local} Hubble constant, $H_0=67\pm 5\kms\mpc^{-1}$, which is consistent with that derived on cosmological scales and does not show evidence for a local `super-Hubble' flow. 

Overall we find that studying the dynamics of nearby galaxies in a broader cosmological context gives us clues on the hierarchical formation of the Local Group constituents as well as meaningful constraints on key cosmological parameters.

\begin{figure*}
\vspace*{0cm}\includegraphics[width=174mm]{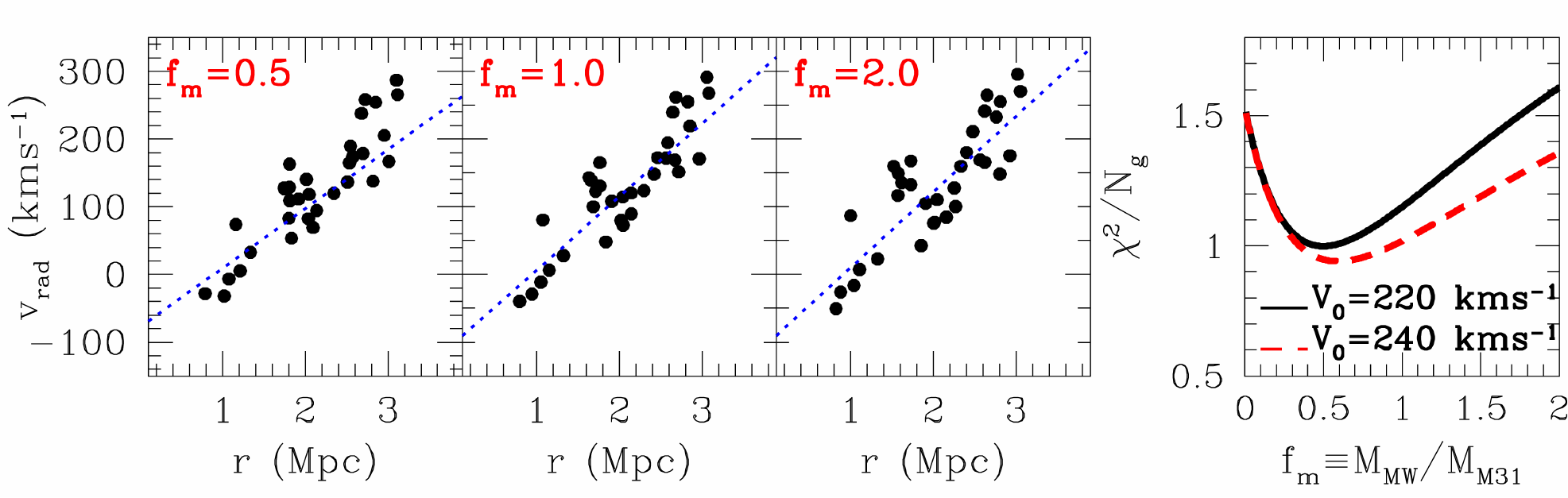}
\caption{Distribution of radii and velocities of the galaxies in Table~\ref{tab:priors} for different values of Milky Way to M31 mass ratio ($f_{\rm m}$). From left to right blue dotted lines show best-fitting straight-lines with $a=-78, -101, -101 \kms$; and $b=87.4, 108.2, 111.7 \kms {\rm Mpc}^{-1}$, respectively.
The fourth panel on the right shows the reduced $\chi^2$ value as a function of $f_{\rm m}$. Notice that models in which the M31 is more massive than the Milky Way significantly increase the scatter of the distribution.}
\label{fig:scatter}
\end{figure*}

\section{Acknowledgements}
We are indebted to Jonathan Diaz who helped us to spot an error in the apex calculation that led to biased constraints on the mass ratio between Andromeda and the Milky Way. We thank Chervin Laporte and Alex Mead for contrasting our numerical set-up against the results of self-consistent cosmological N-body simulations. We also thank the anonymous referee for his/her very useful comments. 
MGW is supported by NSF grant AST-1313045.

\section{Appendix A: Coordinate frame conversion}\label{sec:bary}
In order to calculate the velocity of galaxy tracers with respect to the Local Group we follow Karachentsev \& Makarov (1996) method. First, using Equation~(\ref{eq:com}) we calculate the solar velocity vector in a LG-centric frame (the so-called {\it apex}) as
\begin{equation}\label{eq:vsolar}
{\bf v}_\odot= {\bf V}_0+ {\bf v}_{\rm LSR} - \frac{v_{\rm A}}{1+f_{\rm m}} \hat {\bf r}_{\rm A};
\end{equation}
where $v_{\rm A}$ is the velocity of M31 in the Galactic Standard of Rest and $f_{\rm m}=M_{\rm G}/M_{\rm A}$ is the Milky Way-to-M31 mass ratio. The quantity ${\bf V}_0$ denotes the rotational velocity vector at the solar radius $R_0$, ${\bf v}_{\rm LSR}$ is the solar motion with respect to the Galactic Standard of Rest (GSR), and $ \hat {\bf r}_{\rm A}$ is the unit vector pointing toward the centre of M31. We choose a right-handed Galactocentric coordinate system wherein the sun is located at $(-R_0,0,0)$ and moves with a velocity ${\bf V}_0=(0,V_0,0)$ with respect to the MW centre. Following Sch{\"o}nrich et al. (2010) the Local Standard of Rest (LSR) vector is fixed to ${\bf v}_{\rm LSR}=(11.1, 12.2,7.2)\kms$. While we find that the current uncertainty in the value of $R_0$ (of the order of a kpc) has no visible impact on our results owing to the large heliocentric distances of the sample galaxies gathered from the literature, the value of $V_0$ does introduce a significant element of uncertainty in our measurements. 

The parameter $v_{\rm A}$ in Equation~(\ref{eq:vsolar}) corresponds to the radial velocity of M31 in the Galactic standard of Rest, that is
\begin{equation}\label{eq:v31}
v_{\rm A}=V_{h,{\rm A}} + ({\bf V}_0+ {\bf v}_{\rm LSR})\cdot \hat {\bf r}_{\rm A};
\end{equation}
with $\hat {\bf r}_{\rm A}=(\cos[l_{\rm A}]\cos[b_{\rm A}], \sin[l_{\rm A}]\cos[b_{\rm A}],\sin[b_{\rm A}])$. Following McConnachie (2012) we adopt the following Galactocentric coordinates for M31: $(l,b)_{\rm A}=(-121.2^\circ, -21.6^\circ)$, $d=0.783\mpc$ and $V_{h,{\rm A}}=-300\kms$.

For a given solar apex the radial velocity of a tracer galaxy with respect to the LG centre can be calculated as
\begin{equation}\label{eq:vrad}
v=V_{h} + \Delta v;
\end{equation}
where $V_h$ is the heliocentric radial velocity, and $\Delta v$ is the projection of the galaxy position vector onto the solar apex, i.e. $\Delta v={\bf v}_\odot\cdot \hat {\bf r}_g$ and $\hat {\bf r}_g={\bf r}_g/|{\bf r}_g|$.

\section{Appendix B: The mass ratio between our Galaxy and Andromeda}\label{sec:apex}
For a given set of solar parameters the main uncertainty in the determination of the apex vector reduces to the mass ratio between our Galaxy and M31 ($f_{\rm m}$).
In this work we have explored two methods for constraining the value of the mass ratio between our Galaxy and M31 ($f_{\rm m}$). In \S\ref{sec:like} this parameter is implemented in the likelihood function that fits orbits to the location and velocities of Local Group neighbour galaxies.
 
Here we also explore a {\it geometrical} method to constrain both parameters which does not require solutions to the equations of motion~(\ref{eq:kep}), but whose results turn out to be in excellent agreement with the convolved Bayesian fits explored above. A similar approach was followed by Arp (1986) to measure the the circular velocity of the Milky Way at the solar radius ($V_0$), by Karachentsev et al. (2009) to pin down the location of the Local Group barycentre, and by Karachentsev \& Makarov (1996) and Courteau \& van den Bergh (1999) to constrain the solar apex. 

The method rests upon the assumption that that the distribution of peculiar velocity about the Hubble flow is {\it cold} and invariant under spatial translations. Hence, the fact that neighbour galaxies are distributed over a large area of the sky implies that a biased choice of $f_{\rm m}$ and/or $V_0$ must necessarily introduce a scatter in the distribution of radial velocities derived from Equations~(\ref{eq:vsolar}) and~(\ref{eq:vrad}). It follows that the proper choice of these parameters must be that that minimizes the scatter of the distance-velocity relation of Local Group galaxies when expressed in a Local Group-centric coordinate frame.

Fig.~\ref{fig:scatter} illustrates the dependence of the radial phase-space location of the galaxy sample (see \S\ref{sec:obs} for details). From left to right the first three panels adopt $f_{\rm m}=0.5, 1.0$ and 2.0. By eye it is clear that the values $f_{\rm m}\lesssim 1$ yield narrower distributions than $f_{\rm m}>1$. To measure the scatter in these distributions we fit straight lines $y=a +b r$ (dotted lines). The right-most panel shows the reduced $\chi^2$ values as a function of $f_{\rm m}$. Here $\chi^2$ is defined as
\begin{equation}\label{eq:chi2}
\chi^2=\sum_{i=1}^{N_g}\frac{(v_{\rm rad,i}-y_i)^2}{\sigma_i^2};
\end{equation}
where $\sigma_i^2$ is the variance associated with the measurements of distance ($\epsilon_D$) and heliocentric velocity ($\epsilon_v$) for the $i$th galaxy in the sample,
\begin{equation}\label{eq:sig}
\sigma_i^2= \epsilon_{v,i}^2 + (b \epsilon_{D,i})^2 + \sigma_m^2.
\end{equation}
In \S\ref{sec:results} we find that setting the parameter $\sigma_m\approx 38\kms$ accounts for the presence of an intrinsic dispersion in the distance-velocity relation that goes beyond that introduced by observational errors.

The right panel of Fig.~\ref{fig:scatter} shows that choosing Local Group models where M31 is {\it more} massive than our Galaxy leads to strongly scatted distributions of radii and velocities. The scatter minimizes at $f_{\rm m} \approx 0.5$ if we adopt a circular velocity $V_0=220\kms$, and at $f_{\rm m} \approx 0.55$ for $V_0=240\kms$. This measurement is fully consistent with the constraints on $f_{\rm m}$ derived in \S\ref{sec:results}, and suggests that our Galaxy may be a factor $\sim 2$ less massive than M31\footnote{Note that Karachentsev et al. (2009) find $f_{\rm m}\sim 1$ using very similar arguments. The slightly discrepant result may be due to revised measurements of distances and velocities (e.g. Tucana) and/or the addition of newly discovered systems (e.g. Leo P).}.

{}


\begin{thebibliography}{}
\bibitem[Aragon-Calvo et al.(2011)]{2011MNRAS.415L..16A} Aragon-Calvo, 
M.~A., Silk, J., \& Szalay, A.~S.\ 2011, \mnras, 415, L16 

\bibitem[Arp(1986)]{1986A&A...156..207A} Arp, H.\ 1986, \aap, 156, 207 

\bibitem[Bahl 
\& Baumgardt(2014)]{2014MNRAS.438.2916B} Bahl, H., \& Baumgardt, H.\ 2014, \mnras, 438, 2916 

\bibitem[Barber et al.(2013)]{2013arXiv1310.0466B} Barber, C., Starkenburg, 
E., Navarro, J., McConnachie, A., \& Fattahi, A.\ 2013, arXiv:1310.0466 

\bibitem[Baryshev et 
al.(2001)]{2001A&A...378..729B} Baryshev, Y.~V., Chernin, A.~D., \& Teerikorpi, P.\ 2001, \aap, 378, 729 

\bibitem[Battaglia et al.(2005)]{2005MNRAS.364..433B} Battaglia, G., Helmi, 
A., Morrison, H., et al.\ 2005, \mnras, 364, 433 

\bibitem[Battaglia et al.(2006)]{2006MNRAS.370.1055B} Battaglia, G., Helmi, 
A., Morrison, H., et al.\ 2006, \mnras, 370, 1055 

\bibitem[Bellazzini et al.(2013)]{2013arXiv1310.6365B} Bellazzini, M., 
Oosterloo, T., Fraternali, F., \& Beccari, G.\ 2013, arXiv:1310.6365 

\bibitem[Bovy et al.(2009)]{2009ApJ...704.1704B} Bovy, J., Hogg, D.~W., 
\& Rix, H.-W.\ 2009, \apj, 704, 1704 

\bibitem[Bowden et al.(2013)]{2013MNRAS.435..928B} Bowden, A., Evans, 
N.~W., \& Belokurov, V.\ 2013, \mnras, 435, 928 


\bibitem[Boylan-Kolchin et al.(2013)]{2013ApJ...768..140B} Boylan-Kolchin, 
M., Bullock, J.~S., Sohn, S.~T., Besla, G., 
\& van der Marel, R.~P.\ 2013, \apj, 768, 140 

\bibitem[Chapman et al.(2013)]{2013MNRAS.430...37C} Chapman, S.~C., Widrow, 
L., Collins, M.~L.~M., et al.\ 2013, \mnras, 430, 37

\bibitem[Chapman et al.(2006)]{2006ApJ...653..255C} Chapman, S.~C., Ibata, 
R., Lewis, G.~F., et al.\ 2006, \apj, 653, 255 

\bibitem[Chernin et 
al.(2009)]{2009A&A...507.1271C} Chernin, A.~D., Teerikorpi, P., Valtonen, M.~J., et al.\ 2009, \aap, 507, 1271 

\bibitem[Chernin et 
al.(2004)]{2004A&A...415...19C} Chernin, A.~D., Karachentsev, I.~D., Valtonen, M.~J., et al.\ 2004, \aap, 415, 19 


\bibitem[Chiboucas et al.(2013)]{2013AJ....146..126C} Chiboucas, K., 
Jacobs, B.~A., Tully, R.~B., \& Karachentsev, I.~D.\ 2013, \aj, 146, 126 

\bibitem[Corbelli 
\& Salucci(2000)]{2000MNRAS.311..441C} Corbelli, E., \& Salucci, P.\ 2000, \mnras, 311, 441 


\bibitem[Courteau 
\& van den Bergh(1999)]{1999AJ....118..337C} Courteau, S., \& van den Bergh, S.\ 1999, \aj, 118, 337 

\bibitem[Courtois et al.(2012)]{2012ApJ...744...43C} Courtois, H.~M., 
Hoffman, Y., Tully, R.~B., \& Gottl{\"o}ber, S.\ 2012, \apj, 744, 43 

\bibitem[C{\^o}t{\'e} et al.(2000)]{2000ApJ...537L..91C} C{\^o}t{\'e}, P., 
Mateo, M., Sargent, W.~L.~W., \& Olszewski, E.~W.\ 2000, \apjl, 537, L91 

\bibitem[Deason et al.(2011)]{2011MNRAS.415.2607D} Deason, A.~J., McCarthy, 
I.~G., Font, A.~S., et al.\ 2011, \mnras, 415, 2607 


\bibitem[Deason et al.(2012)]{2012MNRAS.424L..44D} Deason, A.~J., 
Belokurov, V., Evans, N.~W., \& An, J.\ 2012, \mnras, 424, L44 

\bibitem[Diaz et al.(2014)]{2014arXiv1405.3662D} Diaz, J.~D., Koposov, 
S.~E., Irwin, M., Belokurov, V., \& Evans, W.\ 2014, arXiv:1405.3662 

\bibitem[D'Onghia et al.(2010)]{2010ApJ...709.1138D} D'Onghia, E., 
Springel, V., Hernquist, L., \& Keres, D.\ 2010, \apj, 709, 1138 

\bibitem[D'Onghia 
\& Lake(2008)]{2008ApJ...686L..61D} D'Onghia, E., \& Lake, G.\ 2008, \apjl, 686, L61 


\bibitem[Efstathiou(2013)]{2013arXiv1311.3461E} Efstathiou, G.\ 2013, 
arXiv:1311.3461 

\bibitem[Fattahi et al.(2013)]{2013MNRAS.431L..73F} Fattahi, A., Navarro, 
J.~F., Starkenburg, E., Barber, C.~R., 
\& McConnachie, A.~W.\ 2013, \mnras, 431, L73 

\bibitem[Feroz 
\& Hobson(2008)]{2008MNRAS.384..449F} Feroz, F., \& Hobson, M.~P.\ 2008, \mnras, 384, 449 

\bibitem[Feroz et al.(2009)]{2009MNRAS.398.1601F} Feroz, F., Hobson, M.~P., 
\& Bridges, M.\ 2009, \mnras, 398, 1601 


\bibitem[Fraternali et 
al.(2009)]{2009A&A...499..121F} Fraternali, F., Tolstoy, E., Irwin, M.~J., \& Cole, A.~A.\ 2009, \aap, 499, 121 

\bibitem[Geehan et al.(2006)]{2006MNRAS.366..996G} Geehan, J.~J., Fardal, 
M.~A., Babul, A., \& Guhathakurta, P.\ 2006, \mnras, 366, 996 

\bibitem[Giovanelli et al.(2013)]{2013AJ....146...15G} Giovanelli, R., 
Haynes, M.~P., Adams, E.~A.~K., et al.\ 2013, \aj, 146, 15 

\bibitem[Gonzalez et al.(2013)]{2013arXiv1312.2587G} Gonzalez, R.~E., 
Kravtsov, A.~V., \& Gnedin, N.~Y.\ 2013, arXiv:1312.2587 

\bibitem[Governato et al.(1997)]{1997NewA....2...91G} Governato, F., Moore, 
B., Cen, R., et al.\ 1997, New Astronomy, 2, 91 

\bibitem[Hobson et al.(2002)]{2002MNRAS.335..377H} Hobson, M.~P., Bridle, 
S.~L., \& Lahav, O.\ 2002, \mnras, 335, 377 

\bibitem[Hoffman et al.(2008)]{2008MNRAS.386..390H} Hoffman, Y., 
Martinez-Vaquero, L.~A., Yepes, G., 
\& Gottl{\"o}ber, S.\ 2008, \mnras, 386, 390 

\bibitem[Ibata et al.(2014)]{2014ApJ...784L...6I} Ibata, R.~A., Ibata, 
N.~G., Lewis, G.~F., et al.\ 2014, \apjl, 784, L6 

\bibitem[Ibata et al.(2013)]{2013Natur.493...62I} Ibata, R.~A., Lewis, 
G.~F., Conn, A.~R., et al.\ 2013, \nat, 493, 62 

\bibitem[Kahn 
\& Woltjer(1959)]{1959ApJ...130..705K} Kahn, F.~D., \& Woltjer, L.\ 1959, \apj, 130, 705 

\bibitem[Karachentsev et al.(2013)]{2013AJ....145..101K} Karachentsev, 
I.~D., Makarov, D.~I., \& Kaisina, E.~I.\ 2013, \aj, 145, 101 

\bibitem[Karachentsev et al.(2009)]{2009MNRAS.393.1265K} Karachentsev, 
I.~D., Kashibadze, O.~G., Makarov, D.~I., 
\& Tully, R.~B.\ 2009, \mnras, 393, 1265 

\bibitem[Karachentsev et al.(2007)]{2007AJ....133..504K} Karachentsev, 
I.~D., Tully, R.~B., Dolphin, A., et al.\ 2007, \aj, 133, 504 

\bibitem[Karachentsev et 
al.(2003)]{2003A&A...408..111K} Karachentsev, I.~D., Sharina, M.~E., Dolphin, A.~E., \& Grebel, E.~K.\ 2003, \aap, 408, 111 

\bibitem[Karachentsev et 
al.(2002)]{2002A&A...389..812K} Karachentsev, I.~D., Sharina, M.~E., Makarov, D.~I., et al.\ 2002, \aap, 389, 812 

\bibitem[Karachentsev et 
al.(2002)]{2002A&A...383..125K} Karachentsev, I.~D., Dolphin, A.~E., Geisler, D., et al.\ 2002, \aap, 383, 125 


\bibitem[Karachentsev 
\& Makarov(1996)]{1996AJ....111..794K} Karachentsev, I.~D., \& Makarov, D.~A.\ 1996, \aj, 111, 794 

\bibitem[Kerr 
\& Lynden-Bell(1986)]{1986MNRAS.221.1023K} Kerr, F.~J., \& Lynden-Bell, D.\ 1986, \mnras, 221, 1023 


\bibitem[Kirby et al.(2012)]{2012ApJ...751...46K} Kirby, E.~N., Cohen, 
J.~G., \& Bellazzini, M.\ 2012, \apj, 751, 46 

\bibitem[Klypin et al.(2011)]{2011ApJ...740..102K} Klypin, A.~A., 
Trujillo-Gomez, S., \& Primack, J.\ 2011, \apj, 740, 102 

\bibitem[Klypin et al.(2002)]{2002ApJ...573..597K} Klypin, A., Zhao, H., 
\& Somerville, R.~S.\ 2002, \apj, 573, 597 

\bibitem[Kroupa et 
al.(2010)]{2010A&A...523A..32K} Kroupa, P., Famaey, B., de Boer, K.~S., et al.\ 2010, \aap, 523, A32 


\bibitem[Li 
\& White(2008)]{2008MNRAS.384.1459L} Li, Y.-S., \& White, S.~D.~M.\ 2008, \mnras, 384, 1459 


\bibitem[Libeskind et al.(2010)]{2010MNRAS.401.1889L} Libeskind, N.~I., 
Yepes, G., Knebe, A., et al.\ 2010, \mnras, 401, 1889 

\bibitem[Lovell et al.(2011)]{2011MNRAS.413.3013L} Lovell, M.~R., Eke, 
V.~R., Frenk, C.~S., \& Jenkins, A.\ 2011, \mnras, 413, 3013 

\bibitem[Lynden-Bell(1981)]{1981Obs...101..111L} Lynden-Bell, D.\ 1981, The 
Observatory, 101, 111 

\bibitem[Lynden-Bell(1982)]{1982Obs...102..202L} Lynden-Bell, D.\ 1982, The 
Observatory, 102, 202 

\bibitem[Navarro et al.(1997)]{1997ApJ...490..493N} Navarro, J.~F., Frenk, 
C.~S., \& White, S.~D.~M.\ 1997, \apj, 490, 493 


\bibitem[Ma et al.(2013)]{2013ApJ...771..137M} Ma, Y.-Z., Hinshaw, G., 
\& Scott, D.\ 2013, \apj, 771, 137 

\bibitem[Macci{\`o} et al.(2005)]{2005MNRAS.359..941M} Macci{\`o}, A.~V., 
Governato, F., \& Horellou, C.\ 2005, \mnras, 359, 941 

\bibitem[Mackey et al.(2010)]{2010ApJ...717L..11M} Mackey, A.~D., Huxor, 
A.~P., Ferguson, A.~M.~N., et al.\ 2010, \apjl, 717, L11 

\bibitem[Martinez-Vaquero et al.(2009)]{2009MNRAS.397.2070M} 
Martinez-Vaquero, L.~A., Yepes, G., Hoffman, Y., Gottl{\"o}ber, S., 
\& Sivan, M.\ 2009, \mnras, 397, 2070 

\bibitem[McConnachie(2012)]{2012AJ....144....4M} McConnachie, A.~W.\ 2012, 
\aj, 144, 4 

\bibitem[McConnachie et al.(2009)]{2009Natur.461...66M} McConnachie, A.~W., 
Irwin, M.~J., Ibata, R.~A., et al.\ 2009, \nat, 461, 66 

\bibitem[McMillan(2011)]{2011MNRAS.414.2446M} McMillan, P.~J.\ 2011, 
\mnras, 414, 2446 

\bibitem[McQuinn et al.(2013)]{2013arXiv1310.0044M} McQuinn, K.~B.~W., 
Skillman, E.~D., Berg, D., et al.\ 2013, arXiv:1310.0044 

\bibitem[Metz et al.(2007)]{2007MNRAS.374.1125M} Metz, M., Kroupa, P., 
\& Jerjen, H.\ 2007, \mnras, 374, 1125 

\bibitem[Milgrom(1983)]{1983ApJ...270..371M} Milgrom, M.\ 1983, \apj, 270, 
371

\bibitem[Mo et al.(1998)]{1998MNRAS.295..319M} Mo, H.~J., Mao, S., 
\& White, S.~D.~M.\ 1998, \mnras, 295, 319 


\bibitem[Mohayaee 
\& Tully(2005)]{2005ApJ...635L.113M} Mohayaee, R., \& Tully, R.~B.\ 2005, \apjl, 635, L113 

\bibitem[Navarro et al.(1997)]{1997ApJ...490..493N} Navarro, J.~F., Frenk, 
C.~S., \& White, S.~D.~M.\ 1997, \apj, 490, 493 

\bibitem[Partridge et al.(2013)]{2013MNRAS.436L..45P} Partridge, C., Lahav, 
O., \& Hoffman, Y.\ 2013, \mnras, 436, L45 

\bibitem[Pawlowski et al.(2013)]{2013MNRAS.tmp.2207P} Pawlowski, M.~S., 
Kroupa, P., \& Jerjen, H.\ 2013, \mnras, 2207 

\bibitem[Peacock(1999)]{1999coph.book.....P} Peacock, J.~A.\ 1999, 
Cosmological Physics, by John A.~Peacock, pp.~704.~ISBN 
052141072X.~Cambridge, UK: Cambridge University Press, January 1999.,  

\bibitem[Peirani 
\& de Freitas Pacheco(2008)]{2008A&A...488..845P} Peirani, S., \& de Freitas Pacheco, J.~A.\ 2008, \aap, 488, 845 

\bibitem[Peirani(2010)]{2010MNRAS.407.1487P} Peirani, S.\ 2010, \mnras, 
407, 1487 

\bibitem[Pe{\~n}arrubia(2013)]{2013MNRAS.433.2576P} Pe{\~n}arrubia, J.\ 
2013, \mnras, 433, 2576 

\bibitem[Pe{\~n}arrubia et al.(2012)]{2012ApJ...760....2P} Pe{\~n}arrubia, 
J., Koposov, S.~E., \& Walker, M.~G.\ 2012, \apj, 760, 2 

\bibitem[Pe{\~n}arrubia et al.(2010)]{2010MNRAS.406.1290P} Pe{\~n}arrubia, 
J., Benson, A.~J., Walker, M.~G., et al.\ 2010, \mnras, 406, 1290 

\bibitem[Planck Collaboration(2013)]{2013arXiv1311.1657P} Planck 
Collaboration 2013, arXiv:1311.1657 

\bibitem[Rauzy 
\& Gurzadyan(1998)]{1998MNRAS.298..114R} Rauzy, S., \& Gurzadyan, V.~G.\ 1998, \mnras, 298, 114 

\bibitem[Reid et al.(2009)]{2009ApJ...700..137R} Reid, M.~J., Menten, 
K.~M., Zheng, X.~W., et al.\ 2009, \apj, 700, 137 

\bibitem[Sakamoto et 
al.(2003)]{2003A&A...397..899S} Sakamoto, T., Chiba, M., \& Beers, T.~C.\ 2003, \aap, 397, 899 

\bibitem[Sandage(1986)]{1986ApJ...307....1S} Sandage, A.\ 1986, \apj, 307, 
1

\bibitem[Sch{\"o}nrich(2012)]{2012MNRAS.427..274S} Sch{\"o}nrich, R.\ 2012, 
\mnras, 427, 274 

\bibitem[Sch{\"o}nrich et al.(2010)]{2010MNRAS.403.1829S} Sch{\"o}nrich, 
R., Binney, J., \& Dehnen, W.\ 2010, \mnras, 403, 1829 



\bibitem[Seigar et al.(2008)]{2008MNRAS.389.1911S} Seigar, M.~S., Barth, 
A.~J., \& Bullock, J.~S.\ 2008, \mnras, 389, 1911 


\bibitem[Skilling(2004)]{2004AIPC..735..395S} Skilling, J.\ 2004, American 
Institute of Physics Conference Series, 735, 395 

\bibitem[Shaya 
\& Tully(2013)]{2013MNRAS.436.2096S} Shaya, E.~J., \& Tully, R.~B.\ 2013, \mnras, 436, 2096 


\bibitem[Smith et al.(2007)]{2007MNRAS.379..755S} Smith, M.~C., Ruchti, 
G.~R., Helmi, A., et al.\ 2007, \mnras, 379, 755 

\bibitem[Sorce et al.(2013)]{2013ApJ...765...94S} Sorce, J.~G., Courtois, 
H.~M., Tully, R.~B., et al.\ 2013, \apj, 765, 94 

\bibitem[Springel et al.(2008)]{2008MNRAS.391.1685S} Springel, V., Wang, 
J., Vogelsberger, M., et al.\ 2008, \mnras, 391, 1685 


\bibitem[Steinmetz et al.(2006)]{2006AJ....132.1645S} Steinmetz, M., 
Zwitter, T., Siebert, A., et al.\ 2006, \aj, 132, 1645 

\bibitem[Teerikorpi et 
al.(2005)]{2005A&A...440..791T} Teerikorpi, P., Chernin, A.~D., \& Baryshev, Y.~V.\ 2005, \aap, 440, 791 


\bibitem[Tully 
\& Fisher(1977)]{1977A&A....54..661T} Tully, R.~B., \& Fisher, J.~R.\ 1977, \aap, 54, 661 

\bibitem[Tully 
\& Fisher(1987)]{1987JBAA...98...48T} Tully, R.~B., \& Fisher, J.~R.\ 1987, Journal of the British Astronomical Association, 98, 48 


\bibitem[Tully et al.(2006)]{2006AJ....132..729T} Tully, R.~B., Rizzi, L., 
Dolphin, A.~E., et al.\ 2006, \aj, 132, 729 

\bibitem[Tully et al.(2013)]{2013AJ....146...86T} Tully, R.~B., Courtois, 
H.~M., Dolphin, A.~E., et al.\ 2013, \aj, 146, 86 

\bibitem[Tully(2013)]{2013Natur.493...31T} Tully, R.~B.\ 2013, \nat, 493, 
31 

\bibitem[van der Marel et al.(2012)]{2012ApJ...753....8V} van der Marel, 
R.~P., Fardal, M., Besla, G., et al.\ 2012a, \apj, 753, 8 

\bibitem[van der Marel et al.(2012)]{2012ApJ...753....9V} van der Marel, 
R.~P., Besla, G., Cox, T.~J., Sohn, S.~T., 
\& Anderson, J.\ 2012b, \apj, 753, 9 

\bibitem[Veljanoski et al.(2013)]{2013ApJ...768L..33V} Veljanoski, J., 
Ferguson, A.~M.~N., Mackey, A.~D., et al.\ 2013, \apjl, 768, L33 

\bibitem[Vera-Ciro et al.(2011)]{2011MNRAS.416.1377V} Vera-Ciro, C.~A., 
Sales, L.~V., Helmi, A., et al.\ 2011, \mnras, 416, 1377 

\bibitem[Watkins et al.(2010)]{2010MNRAS.406..264W} Watkins, L.~L., Evans, 
N.~W., \& An, J.~H.\ 2010, \mnras, 406, 264 

\bibitem[Wojtak et al.(2013)]{2013arXiv1312.0276W} Wojtak, R., Knebe, A., 
Watson, W.~A., et al.\ 2013, arXiv:1312.0276 

\bibitem[Yepes et al.(2013)]{2013arXiv1312.0105Y} Yepes, G., Gottloeber, 
S., \& Hoffman, Y.\ 2013, arXiv:1312.0105 


\bibitem[Xue et al.(2008)]{2008ApJ...684.1143X} Xue, X.~X., Rix, H.~W., 
Zhao, G., et al.\ 2008, \apj, 684, 1143 

\bibitem[Zhao et 
al.(2013)]{2013A&A...557L...3Z} Zhao, H., Famaey, B., L{\"u}ghausen, F., \& Kroupa, P.\ 2013, \aap, 557, L3 

\bibitem[Zwitter et al.(2008)]{2008AJ....136..421Z} Zwitter, T., Siebert, 
A., Munari, U., et al.\ 2008, \aj, 136, 421 


\end{thebibliography}
\end{document}